\documentclass[reprint,groupedaddress,superscriptaddress,floatfix,longbibliography,tightenlines,nobibnotes,nofootinbib,aps]{revtex4-2}
\usepackage[utf8x]{inputenc}
\usepackage[english]{babel}
\usepackage[T1]{fontenc}
\usepackage{lmodern}
\usepackage{dsfont}

\usepackage[margin=0.6in]{geometry}
\usepackage{fancyhdr}
\usepackage{amsmath,amsfonts,amssymb,amsthm,bm,times}
\usepackage[colorlinks={true}, citecolor={blue}, filecolor={blue}, linkcolor={blue}, urlcolor={blue}]{hyperref}
\usepackage{graphicx,color}
\usepackage[caption=false]{subfig}
\usepackage{textcomp}
\usepackage{lipsum}
\usepackage{verbatim}
\usepackage{blindtext}
\usepackage{natbib}
\usepackage{makecell}
\usepackage{ulem}
\usepackage{booktabs}
\usepackage[toc,page]{appendix}
\usepackage{microtype}

\usepackage{braket}

\begin{document}
	
\title{Valley-dependent 
  transport  through 
graphene quantum dots   \\ due to 
proximity-induced,
staggered  spin-orbit couplings}
	
\author{A. Belayadi}
\email{abelayadi@usthb.dz}
\affiliation{Dept. of Theoretical Physics, University of Science and Technology Houari Boumediene, Bab-Ezzouar 16111, Algeria.}
\affiliation{École Supérieure des Sciences de l'Aliment et Industries Alimentaires, ESSAIA, El Harrach 16200, Algeria.}

\author{P. Vasilopoulos }
\email{p.vasilopoulos@concordia.ca}
\affiliation{Dept. of Physics, Concordia University, 7141 Sherbrooke Ouest, Montr\'eal, Qu\'ebec H4B 1R6, Canada}

\author{N. Sandler}
\email{sandler@ohio.edu}
\affiliation{Dept. of Physics and Astronomy, Ohio University and Nanoscale Quantum Phenomena Institute, Athens, Ohio, USA.}

%%----------------------------------
\begin{abstract}
\noindent We study a system composed of graphene decorated with an array of islands with $C_{3v}$ symmetry that induce quantum dot (IQD) regions via proximity effects and gives rise to several spin-orbit couplings (SOCs). We evaluate transport properties for an array of IQDs and analyze the conditions for realizing isolated valley conductances and valley-state localization. The resulting transmission shows a square-type behavior with wide gaps that can be tuned by adjusting the strength of the staggered intrinsic SOCs. Realistic proximity effects are characterized by weak SOC strengths, and the analysis of our results in this regime shows that the Rashba coupling is the important interaction controlling valley properties. As a consequence, a top gate voltage can be used to tune the valley polarization and switch the valley scattering for positive or negative incident energies. A proper choice of SOC strengths leads to higher localization of valley states around the linear array of IQDs. These systems can be implemented in heterostructures composed of graphene and semiconducting transition-metal dichalcogenides (TMDs) such as MoSe$_2$, WSe$_2$, MoS$_2$, or WS$_2$. In these setups, the magnitudes of induced SOCs depend on the twist angle, and due to broken valley degeneracy, valley polarized currents at the edges can be generated in a controllable manner as well as localized valley states. Our findings suggest an alternative approach for producing valley-polarized currents and propose a corresponding mechanism for valley-dependent electron optics and optoelectronic devices.
\end{abstract}
	
\maketitle
	
%%----------------------------------------------------------------------------------------------------
\section{introduction}\label{sec1}
The study of valleytronic materials is of significant importance in the area of information processing and encoding
\cite{Rohling, Schaibley, Szechenyi} due to the alternative degree of freedom furnished by the carrier's valley momentum in 
addition to the conventionally used charge and/or spin properties.

The benchmark material is a two-dimensional (2D) graphene-based structure with Dirac cones at ${\bf K_1=-K}$ and ${\bf K_2=+K}$ valleys, proposed as a strong candidate for future valley-driven computing devices through the manipulation of valley currents \cite{Culcer, Laird, Hadadi}, i.e., by applying external voltages. However, the lack of external probes or contacts
that can select individual valley currents as ferromagnetic contacts separate spins polarized currents in spintronic devices \cite{Benitez, Lua, Bonbien}, remains a primal obstacle to encoding and information processing through the valley index. 

In addition to transport, the material's optoelectronic properties are also used to access the valley degree of freedom
\cite{Xiao2012, Jingshan, Cresti, Zhang}. Several strategies have used the optical response to control, detect, and monitor valley polarization \cite{Krasnok, Sharma, Rasmus,Wu}. Generally, a combination of gates voltages -implemented via scanning tunneling microscopy- and suitable substrate magnetic materials bring out the mechanism that tunes the desired electronic, spintronic, and valleytronic properties \cite{Maksym, Orlof, Wyatt, SYLi} as proposed and later demonstrated by polarization-resolved photoluminescence experiments \cite{Xiao2012, Cao, Zhang, Aktor, Iwasa}.

An alternative approach to induce valley separation involves exploiting confined geometries. Quantum dots (QDs) can produce valley-filtered currents and are important ingredients in modern nanotechnology devices. Typically, confined geometries that induce valley separation are obtained via a wide variety of methods that include electrostatic confinement produced by a scanning tunneling microscopy tip \cite{Wyatt}, strain fields \cite{Nancy1}, bilayer graphene structures with spatially varying broken sublattice symmetry \cite{Stephen1} and isolated regions defined by local broken sublattice symmetry \cite{Aktor}. In all these setups, valley separation is achieved because of the effects of external fields or due to substrate properties that are extremely difficult to design and control with sufficient precision.  As a consequence, the potential of these geometries to induce {\it selective valley filtering and confinement in a controllable manner and without external fields} remains untapped. To address this issue, we investigate the properties of a proposed heterostructure that exploits proximity effects and periodic spin-orbit interactions.

Ideally, the most efficient way to induce uniform and large staggered spin-orbit couplings (SOC)s on graphene is via proximity to appropriate substrates that break the sublattice symmetry, thus allowing for a clear distinction of the two pseudospins. Recently, the role of  SOCs in valley separation has been addressed in several works, such as graphene deposited on top of hexagonal boron nitride \cite{Kochan-2017} and graphene/TMD heterostructures \cite{Gmitra-2016, Belayadi-tmd-2023}. These setups possess staggered onsite potentials that give rise to various SOCs via different mechanisms \cite{Aktor}. Interestingly, not all types of SOCs will render valley separation as shown by the sublattice independent intrinsic spin-orbit coupling (ISOC) in the Kane and Mele model \cite{Kane-Mele}, or the Rashba SOC (RSOC) that appears in the presence of external fields. However, other appropriately engineered interactions can break the sublattice symmetry, rendering two main effects that we refer to as (1) the rise of a staggered potential with a concomitant gap opening and (2) the emergence of an ISOC in a staggered form that is sublattice dependent (i.e., sublattice-resolved SOC). In this last case, the spin-valley transport is due to the emergence of a valley-Zeeman type of coupling, defined by the ISOC sign change between sublattices \cite{Gmitra-2015, Gmitra-2016, Wang}. This valley Zeeman effect is of great interest because it induces a giant spin lifetime anisotropy in proximitized graphene \cite{Cummings}. Furthermore, Frank et ${\it al}$ \cite{Frank-2018} showed that in narrow-width cells of zigzag-terminated graphene with a staggered ISOC, pseudo-helical and valley-centered states (without topological protection) are localized along the edges. These results are consistent with the bulk system's topological invariant $Z_2=0$.

In this work, we propose a heterostructure composed of graphene and TMD islands that combines the effects of confinement and SOCs in a controllable manner. The model is inspired by recent experiments reported in Ref. \onlinecite{Ren-2023}, with graphene deposited on top of a TMD island that induces a local region with various SOCs, i.e., an induced quantum dot (IDQ). Our proposal generalizes the experimental setup to a periodic array of islands placed below or deposited on the graphene membrane. The TMD islands preserve the underlying $C_{3v}$ symmetry of graphene and introduce SOCs in the electron dynamics via the proximity effect. We analyze the conditions for selective valley state confinement and the generation of valley currents under applied voltages for a generic model that is later applied to specific material combinations.

The paper is organized as follows. In Sec. \ref{sec2}, we briefly present the model for a system composed of a linear array of induced quantum dots in graphene created by proximity effects and including different emerging SOC terms. In Sec. \ref{sec3}, we present numerical results revealing effective mechanisms for valley filtering and confinement. We apply these results to a series of heterostructures composed of different materials with realistic parameters and analyze the effect of relative twisting between the two materials.  A summary and conclusions follow in Sec. \ref{sec4}.

%%----------------------------------------------------------------------------------------------------
\section{Model and methods}\label{sec2}

\begin{figure}[tp]
\centering
\hspace*{-0.35cm}
\includegraphics[width=9.2cm, height=8.5cm]{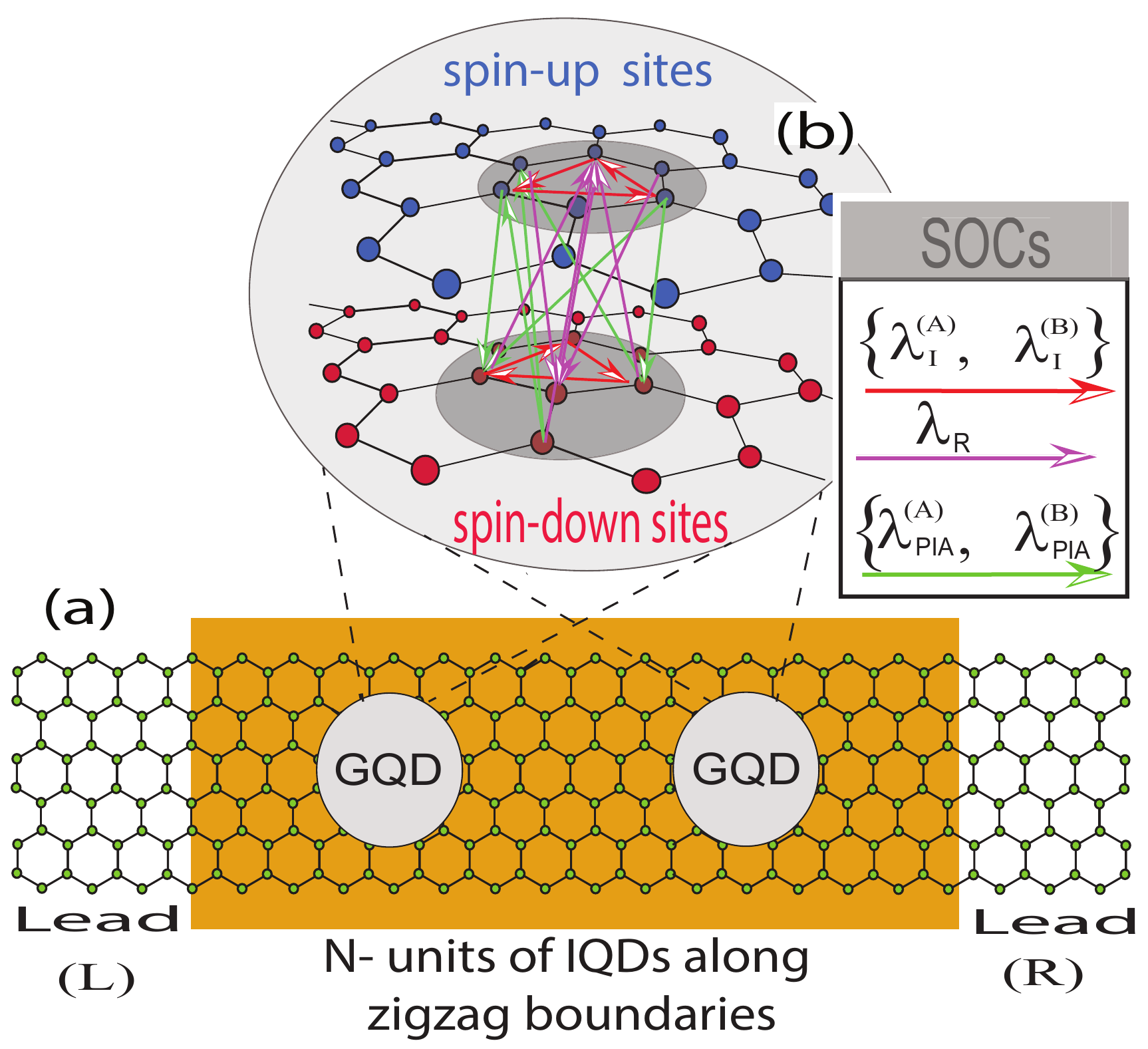}
\vspace{-0.45cm}
\caption{ Panel (a) displays the overall device, which consists of a 2D graphene ribbon with zigzag boundaries decorated with semiconducting transition-metal dichalcogenide (TMD) islands. (b) Zoom-in of a quantum dot made of graphene and TMD. To visualize the emergence of different SOC terms, we duplicate the graphene membrane to emphasize the lift of the spin degeneracy. Each layer corresponds to a different spin component, with the blue (red) membrane representing the spin-up (spin-down) population. The induced SOCs combined with the underlying $C_{3v}$ symmetry give rise to sublattice-resolved intrinsic couplings $ \lambda_{I}^{(A)}, \ \lambda_{I}^{(B)} $, denoted by red arrows, and pseudospin inversion-asymmetric couplings $ \lambda_{PIA}^{(A)}, \ \lambda_{PIA}^{(B)} $ denoted by green arrows. The Rashba $\lambda_R$ coupling is represented by purple arrows. }
	\label{fig1}
\end{figure}

As mentioned above, we propose to study a chain of quantum dots with $C_{3v}$ symmetry in graphene created by proximity effects due to TMD islands. The choice of TMDs that conserve $C_{3v}$ symmetry is made to ensure the largest values of induced spin-dependent couplings in the graphene membrane\cite{Gmitra-2016, Naimer-2021}. Such engineered IQDs will exhibit pseudohelical and valley-centered edge states with potential for device applications \cite{Frank-2018, Hogl-2020}. The salient advantages of such structures are: 1) longer localization lengths for valley states in narrow ribbons, 2) valley Chern numbers and localization lengths independent of RSOCs, and 3) gapless band structures. A schematic picture of the system is shown in Fig. \ref{fig1}.
%%---------------------------- Fig. 1--------------------
%%------------------------- end Fig. 1--------------------
The deposition of adsorbates on graphene, or of graphene membranes on islands, results in profound changes in the electronic structure that depend on the locally conserved symmetries as described in Ref. \cite{Gmitra-2016}. The most important effects are the emergence of i) an effective staggered potential due to the broken reflection symmetry imposed by different orbital interactions experienced by the carbon atoms in proximity to the different atomic species of the TMD material and ii) several sublattice dependent next-nearest neighbor hopping terms originated from the proximity-induced spin-orbit interactions.
In this case, the system can be described by an extension of the models by Kane and Mele \cite{Kane-Mele} and Haldane \cite{Haldane}. The Hamiltonian for the QD regions is  given by: \cite{Kochan-2017, Gmitra-2016, Belayadi-tmd-2023, Ghiasi}
%%%========== Eq.1 ===============
\begin{eqnarray}\label{eq1}
\notag
&H_{QD}&= -t \sum_{\langle i,j\rangle}a_{i s}^\dagger b_{j s}
\notag \\*
&+\sum_{ \left\langle i \right\rangle }&\Delta\left( \xi^{(A)} {\bf a}_{i s}^\dagger {\bf a}_{i s}+ \xi^{(B)} {\bf b}_{i s}^\dagger {\bf b}_{is} \right)
\notag \\*
\hspace{-0.6cm}
\notag
&+\Big( \frac{2i}{3}\Big)& \sum_{\left\langle i, j \right\rangle \sigma, \sigma'}\left( \lambda_{R} \ {\bf a}_{i\sigma}^\dagger {\bf b}_{j\sigma} \right)\left[ {\bf \hat{s}}\otimes {\bf d}_{ij}\right]_{\sigma, \sigma'}\\*
\hspace{-0.6cm}
&+\Big( \frac{i}{3\sqrt{3}} \Big)& \sum_{\left\langle \left\langle i, j \right\rangle \right\rangle \sigma}^{}\nu_{ij}\left( \lambda_{I}^{(A)} {\bf a}_{i\sigma}^\dagger {\bf a}_{j\sigma}+\lambda_{I}^{(B)} {\bf b}_{i\sigma}^\dagger {\bf b}_{j\sigma} \right)\left[ {\bf \hat{s}}_{z}\right]_{\sigma,\sigma}
\end{eqnarray}
where $t$ is the nearest neighbor hopping between sites $i$ and $j$ (note that these are spin-preserving processes). $\Delta$ is the
 staggered potential induced by the TMD islands in the dot region. This potential is sublattice-dependent with 
  $\xi^{(A)} =1$ ($\xi^{(B)} =-1$), rendering opposite signs for the induced gaps, i.e., $\Delta^{(A)} = -\Delta^{(B)} = \Delta$ for sublattice A (B). The Rashba interaction (RSOC) is expressed in terms of the coupling $\lambda_R$. This coupling breaks the $z$ inversion symmetry while exchanging the spin of different sublattices. Here, the $\textbf{d}_{ij} $  vector  connects site $j$ to $i$. The terms $\lambda_I^{(A)}$ and $\lambda_I^{(B)}$ represent the intrinsic SOCs (ISOC) between next-nearest neighbors. These terms connect the same sublattices and spins in the clockwise ($\nu_{ij}=-1$) or anticlockwise ($\nu_{ij}=-1$) direction from site $j$ to site $i$. Finally, the spin is denoted by the vector $\widehat{\textbf{s}}$ with components written in terms of Pauli matrices. It is worth restating that the SOCs exist only within the QD regions; outside the system is described by the Hamiltonian of pristine graphene.

Working with a Hamiltonian in reciprocal space is more convenient for studying valley properties. The resulting effective Hamiltonian is obtained by linearizing Eq. (\ref{eq1}) around the ${\bf K_1}$ and ${\bf K_2}$ valleys labeled below by the valley index $\kappa=- 1$ and $\kappa=+ 1$, respectively. The final expression is given in the form $H_{QD}=H_k+H_{\Delta}+H_R+H_I$ \cite{Frank-2018}, where:
\begin{eqnarray}
\label{eq2}
H_k&=&\hbar v_F \left( \kappa k_x \sigma_x + k_y \sigma_y \right)s_0,\\*	
\label{eq3}
H_{\Delta}&=& \Delta \sigma_zs_0,\\*
\label{eq4}
H_R&=&\lambda_R \left( -\kappa \sigma_x s_y + \sigma_y s_x \right)s_0,\\*
\label{eq5}
H_I&=&(\kappa/2) 
\big[ \lambda_I^{(A)}\left( \sigma_z+\sigma_0 \right) + \lambda_I^{(B)} \left( \sigma_z-\sigma_0 \right)\big] s_z.
\end{eqnarray}

The Fermi velocity $v_F$ is expressed in terms of the hopping $t$ as $ v_F=\sqrt{3}a_{0}t/2\hbar$ where $a_{0}$ is the lattice constant. The pseudospin is denoted by the Pauli matrices $\sigma$, and $s_0$ denotes the spin identity matrix. 

While Eqs.(\ref{eq2}-\ref{eq5}) provide an intuitive picture of the effect of each SOC term on the valleys, the results presented in the following sections are obtained by combining the S-matrix formalism with the tight-binding model for a zigzag terminated ribbon. This boundary condition preserves the valley quantum number and the valley topological properties of graphene. We compute the valley-polarized conductance with the Landauer-B\"{u}ttiker approach: 
\begin{equation}\label{eq21}
G_{\kappa}^{n,m} = (e^2/h) \left| S_{\kappa}^{n,m} \right|^2 , \qquad (n, \ m \equiv L, R).
\end{equation}
Here $S_{\kappa}^{n,m}$ is the scattering matrix element between left (L) and right (R) leads for a given valley index ${\kappa}=\pm 1$ \cite {Datta-1995}. Thus, our calculations exploit the formalism with valley-dependent local currents as defined in Ref. \cite{kwant}. Details regarding the computation of the valley conductance and currents are presented in Appendices \ref{app-A} and \ref{app-B}.

\section{ Results }\label{sec3}
In this section, we present numerical results for the valley-polarized conductance for a range of structures that contain from a single IQD  ($n=1$) to a chain ($n>1$) of IQDs. To emphasize the qualitative features resulting from the competition among the different interactions in the model, we adjust the parameters' values accordingly to present the main findings. This procedure is usually applied to identify the role played by the various interactions\cite{Garcia-2016, Frank-2018, Cysne-2018}. We note, however, that for accurate setups, one expects weaker values for SOCs from proximity effects, and we address this situation in Sec.\ref{sub-b}.

\subsection{\bf Qualitative analysis of results}\label{sub-a}

Following Refs.\cite{Kane-Mele} and \cite{Frank-2018}, we solve the model using the following ranges for the various SOC parameters: $\lambda_R \le 0.075t$ and $\lambda_I^{(A)}=-\lambda_I^{(B)} \le  \sqrt{27}0.06t$ in units of $t=1$ \cite{Frank-2018}.

We consider  $n$  symmetric quantum dots ($n=1, 2, 3, 4$), with the same spin-orbit parameters, arranged in a chain with the same radius $r_0=7$ nm and the space in-between them set to $d=5r_0$. The circular geometry of the dots is chosen to preserve the symmetries of the original model and provide maximum localization of valley states, as discussed below. We set the incident energy at $E_F = 0.035t$ and consider a ribbon width $w=30$ nm, with zigzag boundaries. For this range of ISOC values, the valley Zeeman effect is the most important term as the topological invariant $Z_2$ remains trivial, i.e.,  $Z_2 = 0$ and the bulk system is in the same topological phase as long as the ISOC is staggered, i.e., $\lambda_I^{(A)}=-\lambda_I^{(B)}$ \cite{Frank-2018, Hogl-2020}. 

\begin{figure}[tp]
	\centering
	\includegraphics[width=9cm, height=4cm]{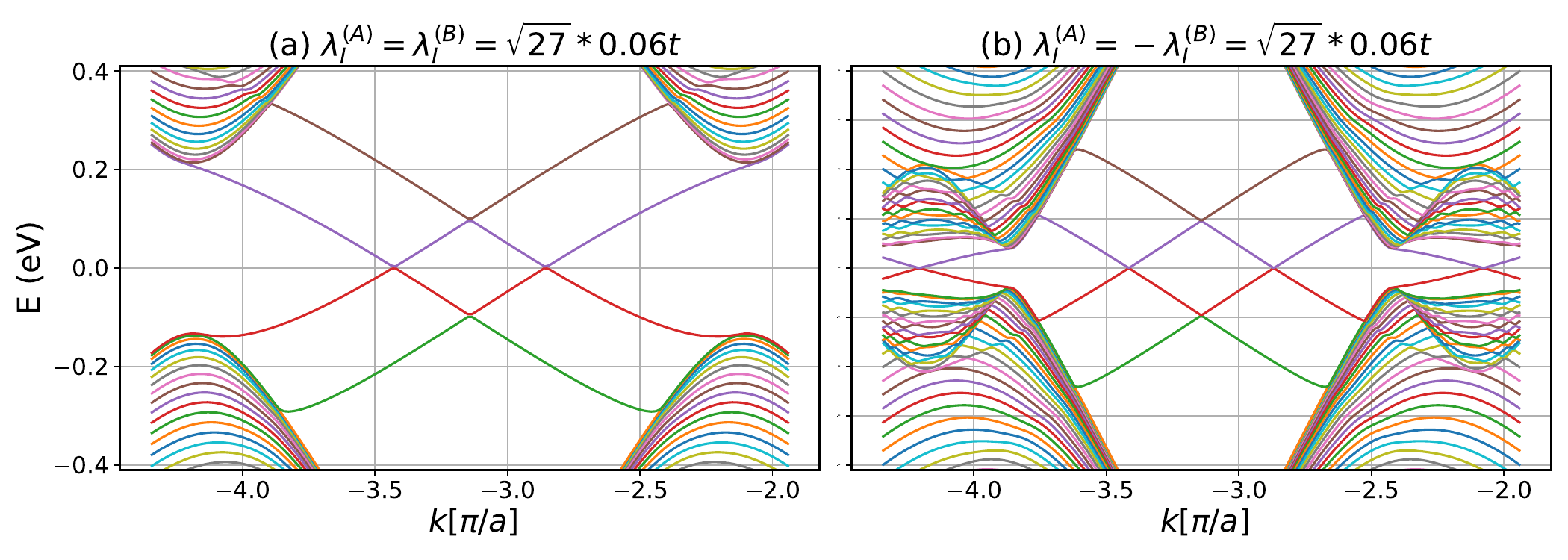}
	\includegraphics[width=9cm, height=4cm]{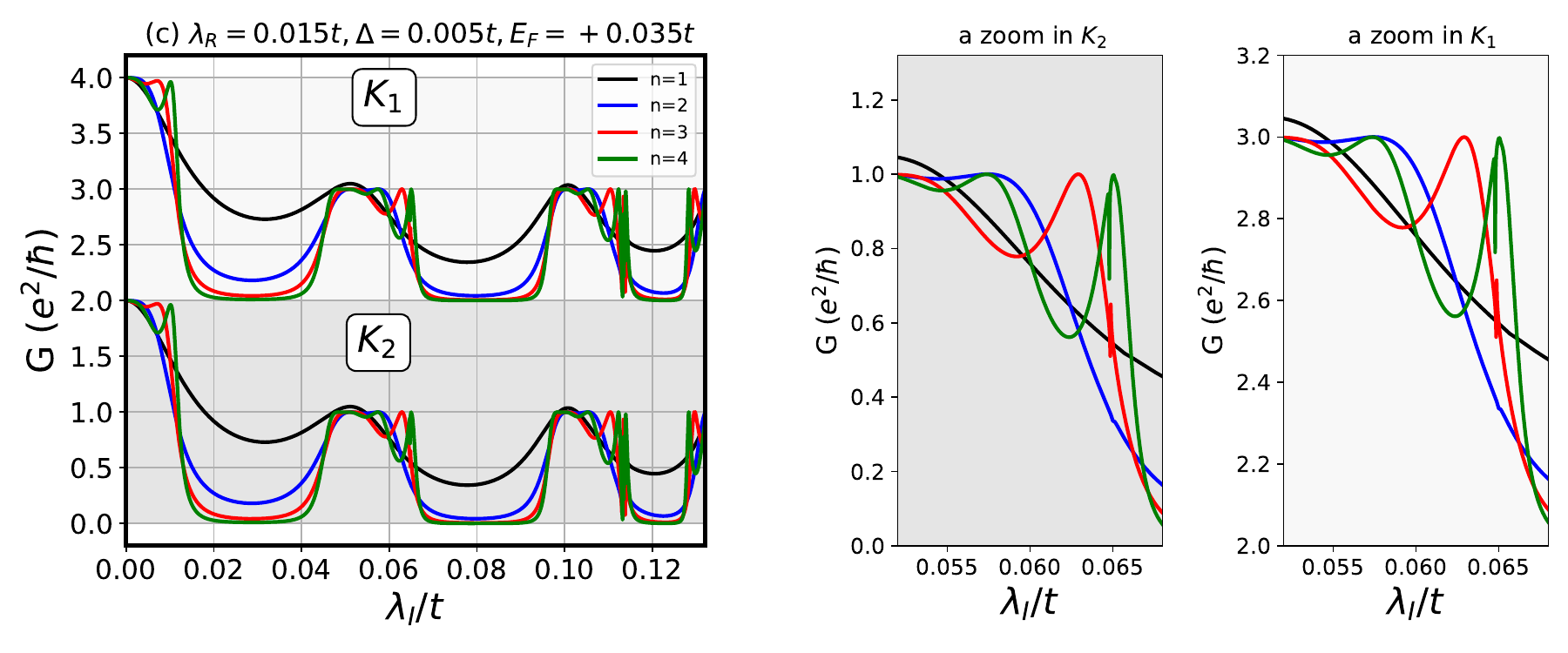}
	\includegraphics[width=9cm, height=4cm]{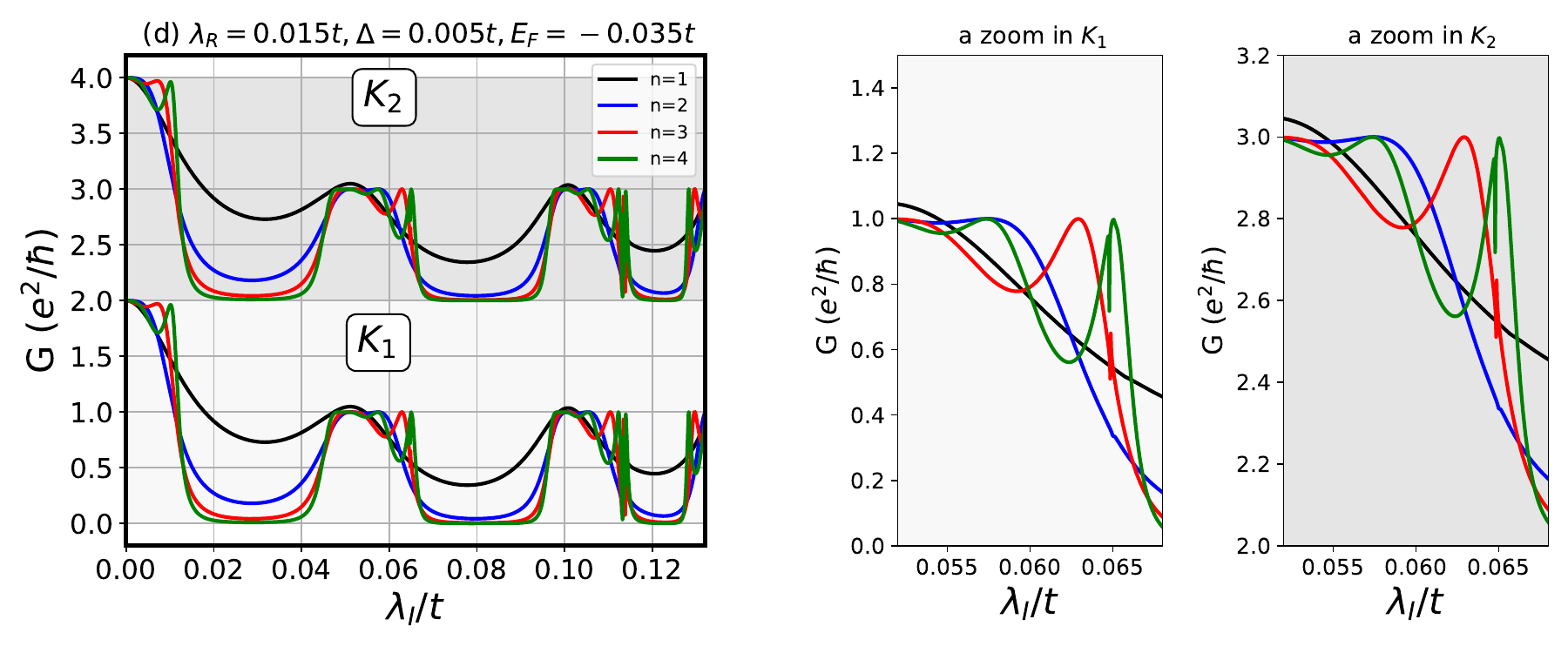}
	\includegraphics[width=9cm, height=4cm]{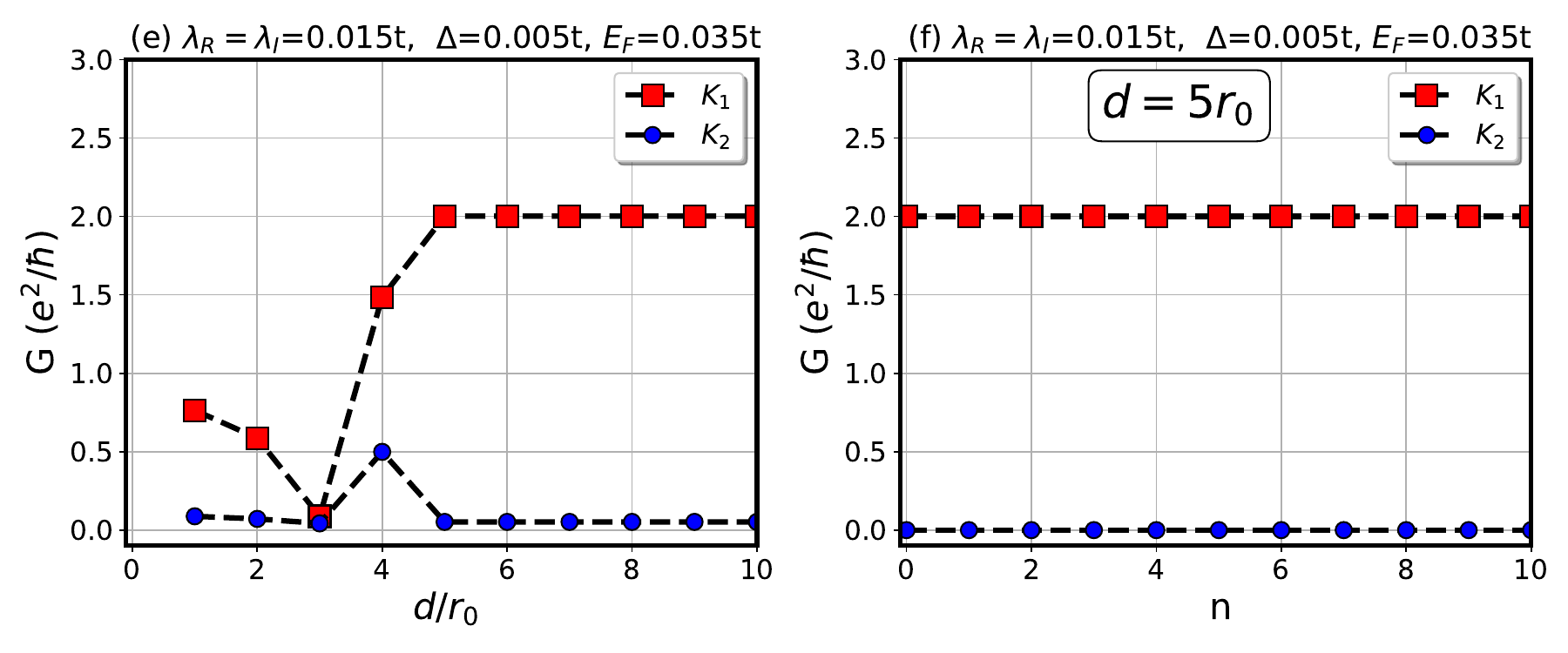}
	\vspace{-0.65cm}
	\caption{Energy bands for uniform $\lambda_I^{(A)}=\lambda_I^{(B)}$ (a), and staggered $\lambda_I^{(A)}=-\lambda_I^{(B)}$) (b) for a 30 nm wide zigzag ribbon, $\Delta=0.1 t$ and $\lambda_R=0.075 t$. Panels (c) and (d) show valley conductances vs intrinsic spin-orbit length $\lambda_I$ for several IQDs with staggered ISOC. Narrow panels on the right correspond to zoom-ins for each valley as a function of $\lambda_I$. Panels (e) and (f) correspond to conductance vs inter-dot spacing $d$ and number $n$ of IQDs.}
	\label{fig2}
\end{figure}

The reduced number of group symmetries of pristine graphene due to the proximity of the TMD island is reflected by the large number of lower-symmetry allowed SOC parameters. The effects of these couplings are observed in Fig, \ref{fig2} (a, b), where the band structure near the two Dirac cones appears clearly modified. Fig. \ref{fig2} (a) shows results in the absence (a) and presence (b) of a staggered potential. The changes include a newly open gap and additional edge bands resulting from the ISOC. These linear bands represent edge states that are unique to the staggered case $\lambda_I^{(A)} = -\lambda_I^{(B)}$ while they disappear in the uniform regime $\lambda_I^{(A)}=\lambda_I^{(B)}$. Similar results have been found in Ref.\cite{Frank-2018} where more details about the valley selection and its related sublattice occupation are discussed. Based on these findings, a proximity effect that induces a staggered ISOC might lead to the emergence of valley currents and the localization of valley-centered sublattice polarized edge states. Indeed, these results manifest themselves in transport, providing valley states as conducting channels, as shown in Fig. \ref{fig2}(b), an issue discussed in the coming subsections.

%%---------------------------- Fig. 2--------------------
%%------------------------- end Fig. 2--------------------

\subsubsection{Valley-dependent conductance through proximity IQDs}\label{sub-a-1} 

Figure \ref{fig2}(c) and (d) display the conductance through a group of $n$ symmetric quantum dots with staggered ISOC.  
One interesting observation is the approximately square-type dependence revealing wide gaps that can be made more pronounced by changing the strength of ISOC for $n \ge 3$ QDs. We observe that $100 \%$ ($0 \%$) of the detected conductance results from the flow of electrons through the  $\kappa=-1$ ($\kappa= +1$) valley for positive incident energy ($E_F = 0.035t$), while $0 \%$ ($100 \%$) occurs for negative incident energy ($E_F=-0.035t$). The opposite behavior is obtained for the complementary valley. Interestingly, the gaps occur at different ranges of ISOC values, with the emerging valley-polarized current being switched from one valley to the other within the gap region by an appropriate change of $E_F$. Furthermore, we observe a decrease in the widths of the gaps with increasing SOC strength, which suggests they might vanish for high enough values of ISOC.

An analysis of Figs. \ref{fig2} (c) and (d) reveal resonances in the transmittance at around the value $\lambda_I=0.065t$, with the spectrum in the zoom-in panels showing that the IQDs confine electrons with index $\kappa=+1$ ($\kappa=-1$) at positive (negative) energy. Hence, the scattering through the IQD region tends to zero accordingly. We observe that the number and sharpness of the resonance weakly depend on the number of IQDs along the chain, as seen for $n=1, 2, 3$ curves that present at least one resonant state each at similar values of ISOCs. More details about the observed confinement are addressed below in Sec.\ref{sub-b-3}, and we discuss these results, including realistic parameters, in \ref{sub-c}.

Additionally, the conductance response shows several interesting characteristics: (1) It exhibits an oscillating behavior that becomes more pronounced with increasing $n$. In this case, the conductance oscillations arise from mode mixing, and their number depends on the number $n$ of the IQDs, as shown in the zoom-in of Fig. \ref{fig2} (c) and (d). (2) The conductance plateaux become better defined as $n$ 
increases ($n \ge 3$), with values $G(\kappa=+1)= 0$ and $G(\kappa=-1)= 2G_0$ for panel (c) and $G(\kappa=+1)= 2G_0$ and $G(\kappa=-1)= 0$ for panel (d). 

These results suggest that the conductance plateaux are due to states that become valley polarized at specific strengths of the staggered ISOC, e.g., in the range $0.015t$ $\le \lambda_I^{(A)}=-\lambda_I^{(B)}\le 0.04t$ as shown in Fig.\ref{fig2} (c). Interestingly, when $E_F$ is negative, the valley polarization is reversed, as shown in panel (d) within the same range of values for $\lambda_I$. Consequently, the transmitted current can be made to be valley polarized from either one of the two Dirac points ${\bf K_1}$ or ${\bf K_2}$ depending on the incident energy.

\subsubsection{Dependence on coherent inter-dot electron transfer}\label{sub-a-2} 

\begin{figure}[tp]
	\centering
	\includegraphics[width=9cm, height=2.2cm]{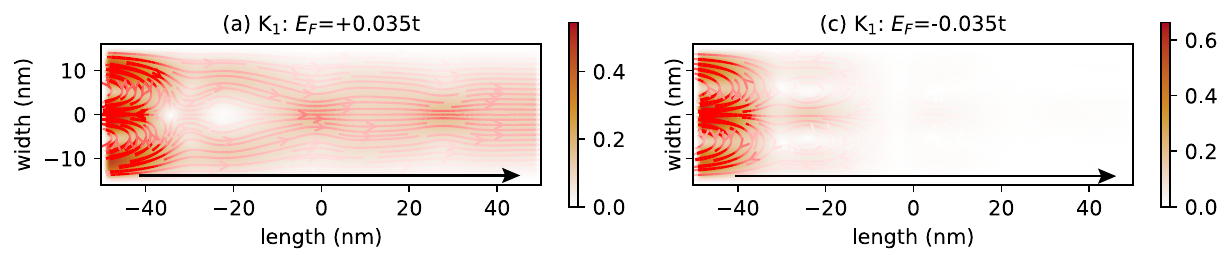}
	\includegraphics[width=9cm, height=2.2cm]{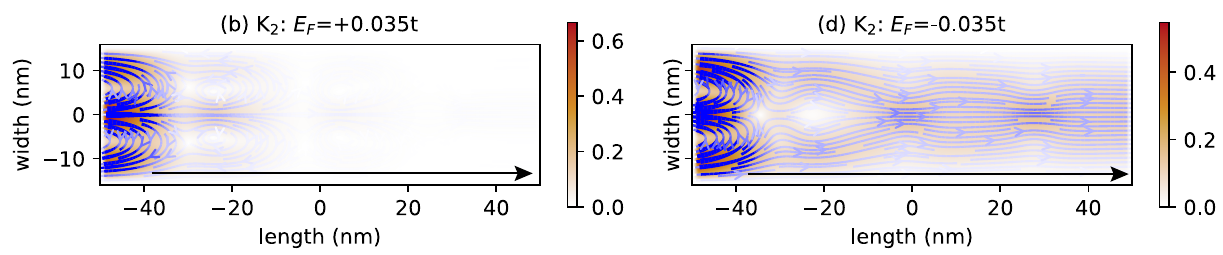}
	\includegraphics[width=9cm, height=2.cm]{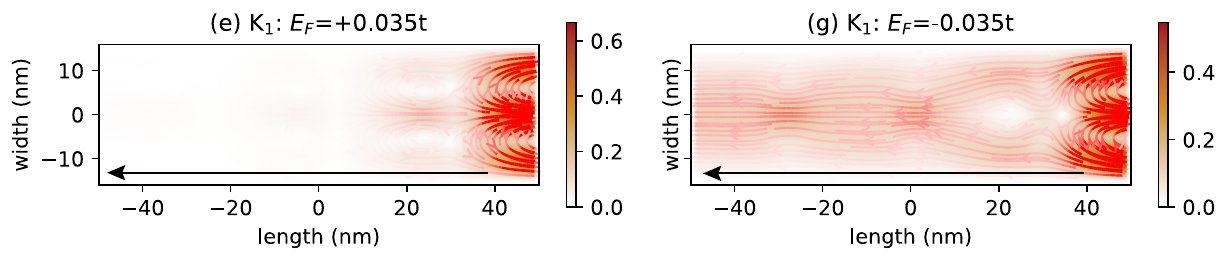}
	\includegraphics[width=9cm, height=2.cm]{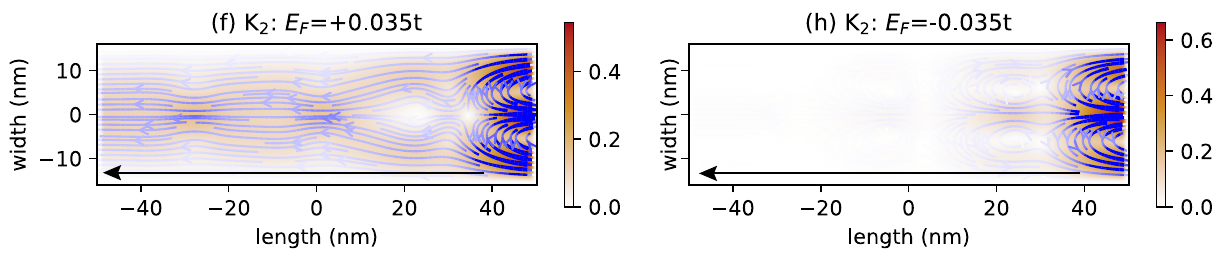}
	\vspace{-0.35cm}
	\caption{Local current mapping, within the gap regions ($\lambda_{I}^{(A)} = -\lambda_{I}^{(B)} = 0.025t$) shown in Fig. \ref{fig2} (c) and (d), for both valleys for a given bias voltage while panels (e) to (h) show results for opposite biases. The left (right) panels are for positive (negative) values of $E_F$. Black arrows indicate the direction of the bias-incident current. Red and blue solid lines indicate the direction of corresponding current flows.}
	\label{fig3}
\end{figure}

Within the conductance gap regions in Fig. \ref{fig2} (c) and (d), and depending on the bias defined by the sign of the Fermi energy, the valley that appears in the output with $T=1$ seems to be barely scattered by the IQDs irrespective of the number of dots in the chain. Inversely, the states from the valley that appear in the output with $T=0$ are strongly reflected by the IQDs even for the shortest chain furnished by only one dot. To better understand these current profiles, we plot in Fig. \ref{fig3} the local valley current through a chain of three IQDs. As the figures show,  the transferred valley current is due to electron scattering processes that involve inter-dot hoppings between neighboring dots, as illustrated in the figure, that display uniform local densities everywhere between the dots for both valleys (Fig. \ref{fig3} (a) and (d)). Notice also that in this case, the conductance is practically the same as for the zigzag terminated graphene ribbon with one or more quantum dots, as depicted in Fig. \ref{fig2} (f). 

%%---------------------------- Fig. 3--------------------
%%------------------------- end Fig. 3--------------------

The above observations raise the question of the inter-dot spacing $ d$'s influence on the valley-transmission output. We have proposed that the valley-scattering processes for complete transmission occur in the presence of effective inter-dot hopping between neighboring IQDs. To test this hypothesis, we analyze the role of $d$ on the valley transmittance. As shown in Fig. \ref{fig2} (e), the inter-dot hopping is present with an inter-dot space $d \ge 5 r_0$. We observe non-stable transmission signatures if $d \le 5 r_0$. In this regime, the edges of the IQDs are closer to each other, and the transmittance peaks show values less (more) than unity (zero) for the ${\bf K_1}$ (${\bf K_2}$) valley. These might be related to mode mixing that impedes the propagation of specific transverse modes due to edge-dot coupling. Interestingly, when the $d$ value is large enough, the valley transmittance becomes stable, represented by steady curves with a uniform local density between the dots. 

\begin{figure}[tp]
	\centering
	%\hspace*{-0.4cm}
	\includegraphics[width=9cm, height=4cm]{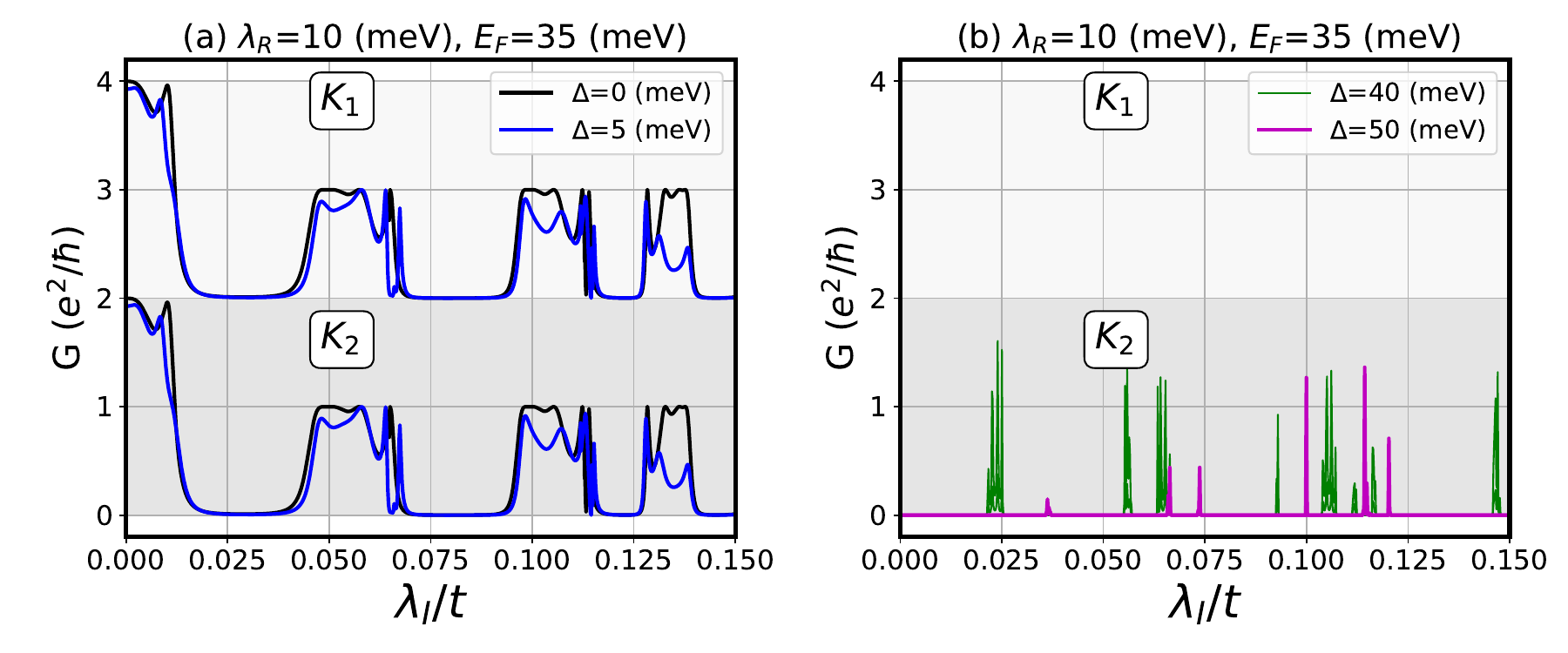}
	\includegraphics[width=9cm, height=4cm]{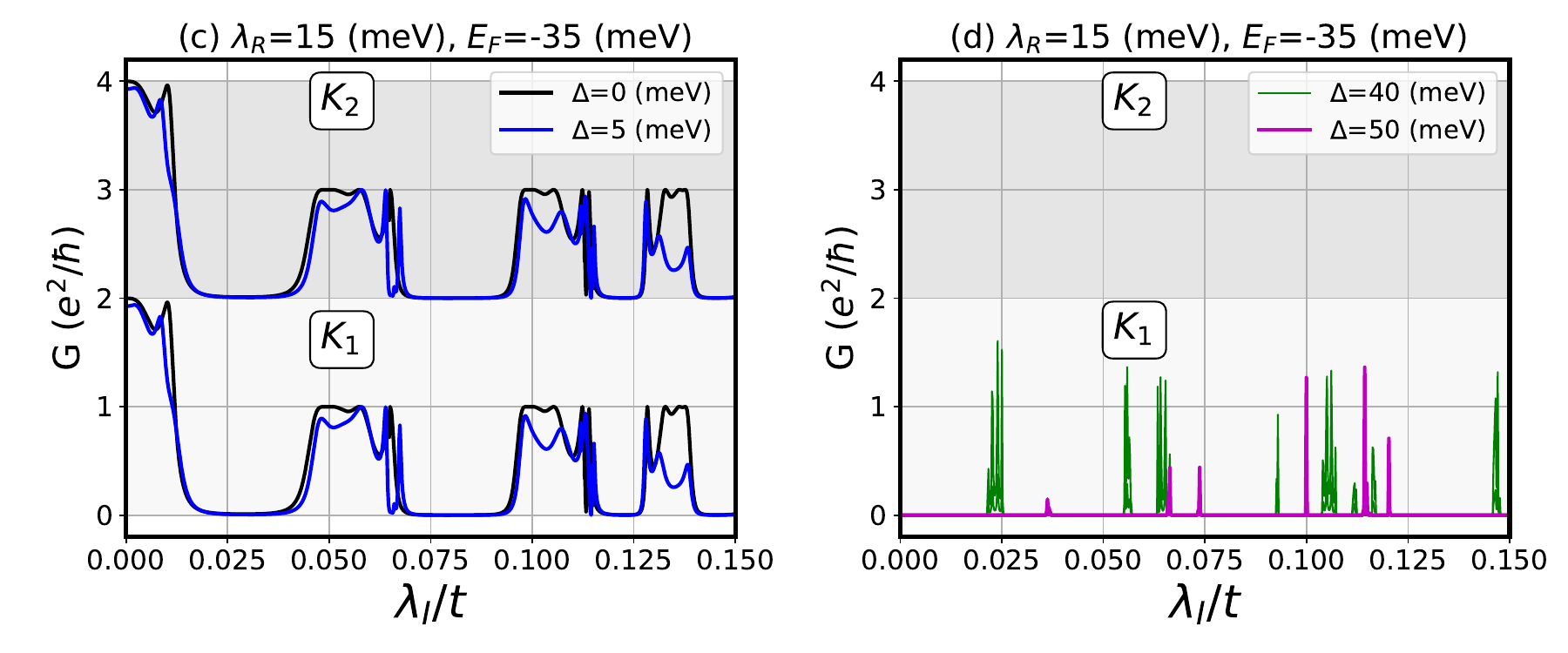}
	\vspace{-0.75cm}
	\caption{Valley conductance vs intrinsic spin-orbit length $\lambda_I$ for different values of staggered potential $\Delta$. The left (right) panels correspond to $\Delta < E_F$ ( $\Delta > E_F$).}
	\label{fig4}
\end{figure}

Finally, we analyze the role of the staggered potential $\Delta$ on the valley-transmission output, as shown in Fig. \ref{fig4}. The staggered potential opens a gap, playing the role of a potential barrier. Consequently, the valley imbalance is maintained for all values of $E_F < \Delta$. However, the valley polarization is rapidly destroyed when $E_F \ge \Delta$. When the confining potential of the IQD is bigger than the incident energy, the current is completely blocked, and only resonant states from each valley ${\bf K_1}$ $({\bf K_2})$ are allowed to transmit current. In this context, the choice of incident energy must consider the strength of the IQD confinement potential.

Fig. \ref{fig4}(b) and (d) show that the conductance is non-zero only for particular resonances where scatterings occur at specific values of the ISOCs and energies suited by edge-dot coupling \cite{Myoung, Chul}. In this context, the heterostructures of graphene and hBN would exhibit this regime since, in this system, the confinement potential $\Delta$ is significant \cite{Zollner-2019} and therefore, its influence on the valley filtering process would have a negative impact.

The results discussed in Fig. \ref{fig4} are essential since they allow us to establish a criterion for defining the best materials for islands. Indeed, for a better response, it is necessary to use a material that provides a weak or zero confinement potential. If so, the IQDs will operate efficiently, and the overall system might be used to monitor valley-driven current by either tuning the ISOC or the RSOC. The question we might ask then is which materials are better suited for this response. Such a case would be materialized in heterostructures of twisted graphene and monolayers of transition-metal dichalcogenides (TMDCs). In such scenarios, the twisting angle substantially decreases the confinement potential, and the dominant parameters will be the valley-Zeeman and RSOCs \cite{Naimer-2021}. A concrete example of a realistic proximity effect is discussed later on.

An alternative description of these effects is by considering the area around the IQDs with wider staggered potentials as an electron/hole bilayer system where the electrons are essentially required to overcome an offset barrier to be scattered through. This action is analogous to the massless-massive electron-hole system in a transverse electric field in graphene on a TMD substrate. In this context, the value of the Rashba coupling has a direct effect since it impacts the offset barrier. Due to its complexity, we postpone the study of these effects for future work. (For more details, see Refs. \cite{Belayadi-tmd-2023} and \cite{Gmitra-2016}.) 

%%---------------------------- Fig. 4--------------------
%%------------------------- end Fig. 4--------------------

\subsubsection{Valley-Hall and bulk conductivities}\label{sub-a-3} 

\begin{figure}[tp]
\centering
\hspace*{-0.4cm}
\includegraphics[width=9cm, height=2.2cm]{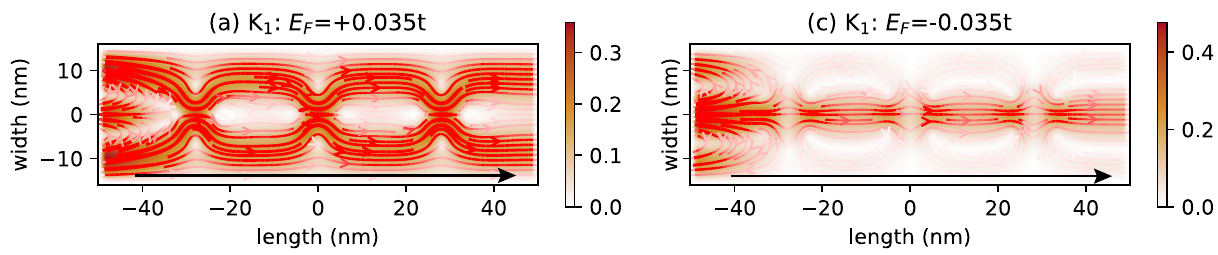}
\includegraphics[width=9cm, height=2.2cm]{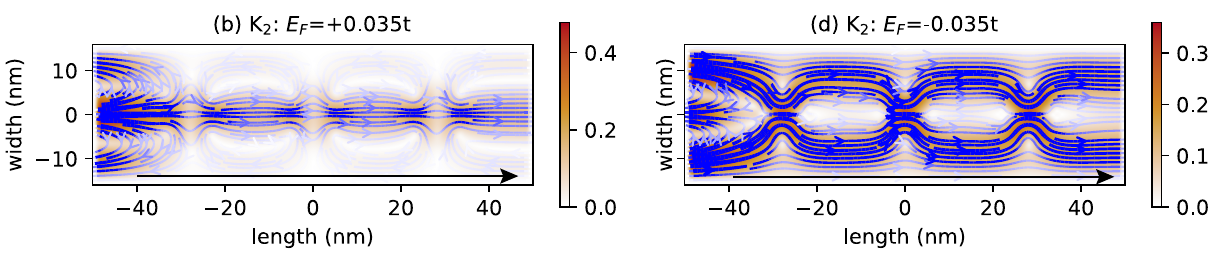}
\includegraphics[width=9cm, height=2.2cm]{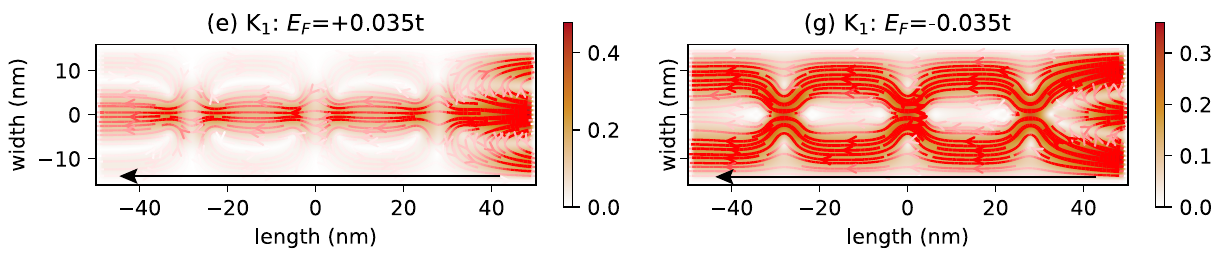}
\includegraphics[width=9cm, height=2.2cm]{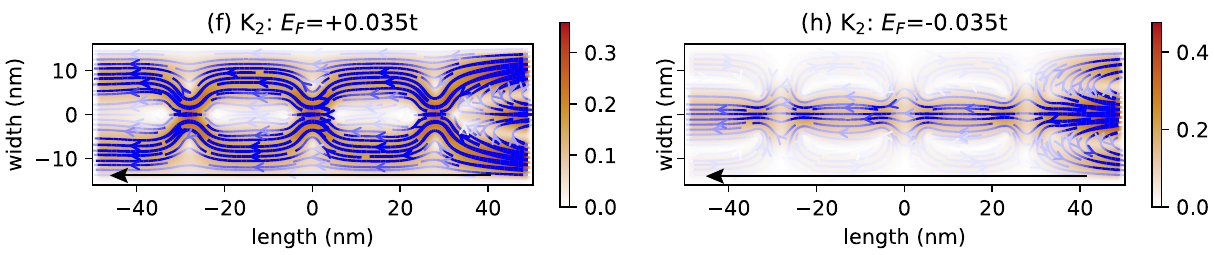}
\vspace{-0.35cm}
\caption{Local valley polarized currents at ${\bf K_1}$ and ${\bf K_2}$ valleys, in the presence of only staggered intrinsic SOC, $\lambda_{I}^{(A)} = - \lambda_{I}^{(B)}=0.075t$. The left (right) panels are for $E_F>0$ ($E_F<0$). }
\label{fig5}
\end{figure}

We briefly discuss the emergence of Hall and bulk conductivities in these structures. We visualize the origin of these conductivities by mapping the local current flow and highlighting the valley polarization with solid blue (red) curves indicating the ${\bf K_1}$ (${\bf K_2}$) index in Fig. \ref{fig5}. Analysis of the figure reveals that the generated valley currents are composed of bulk-driven and  Hall-driven currents. In this process, the local current where both valleys are scattered shows that each valley is conducting with either bulk or Hall currents depending on the sign of the Fermi energy. This analysis suggests that a periodic array of dots with the best choice of SOCs offers an alternative mechanism for generating valley-neutral Hall currents since both valleys contribute to the current in the same direction, although through different regions. A realistic example is discussed in Sec. \ref{sub-c} where we find induced SOC terms that facilitate valley-Hall current for a given TMDs island and twist angle.

%%---------------------------- Fig. 5--------------------
%%------------------------- end Fig. 5--------------------

\subsection{Description of the model with weak couplings}\label{sub-b}
A quantitative description of the model with realistic parameters requires discussing the above results for the weaker strengths of the various SOCs. An important question is how the valley-conductance is affected by changes in the RSOC values.

\subsubsection{Valley-dependent conductance by tuning the Rashba coupling}\label{sub-b-1}

In this context, we consider weaker ISOCs as expected in realistic settings and an increased and controllable strength of RSOC. We notice that IQDs refer, in reality, to any heterostructure that exhibits $C_{3v}$-symmetry. This section considers only ISO and RSO couplings, while the pseudospin inversion asymmetry (PIA) coupling in systems with broken inversion symmetry is discussed in the following section. A typical example of such a setup would be twisted graphene/transition-metal-dichalcogenides heterostructures since the relative rotation leads to a negligible value of PIA coupling. To this end, the IQDs with $C_{3v}$-symmetry might lead to potential applications since the Rashba coupling can be tuned using a transverse electric field. As realized, we consider weak staggered couplings $\lambda_I^{(A)}= -\lambda_I^{(B)}=0.015t$, we inject electrons with $E_F= \pm 0.035t$ into several-identical QDs, and by shifting the  Rashba coupling with the aid of top gates, we compute the valley conductance through the system.

\begin{figure}[tp]
	\centering
	%\hspace*{-0.4cm}
	\includegraphics[width=9cm, height=4cm]{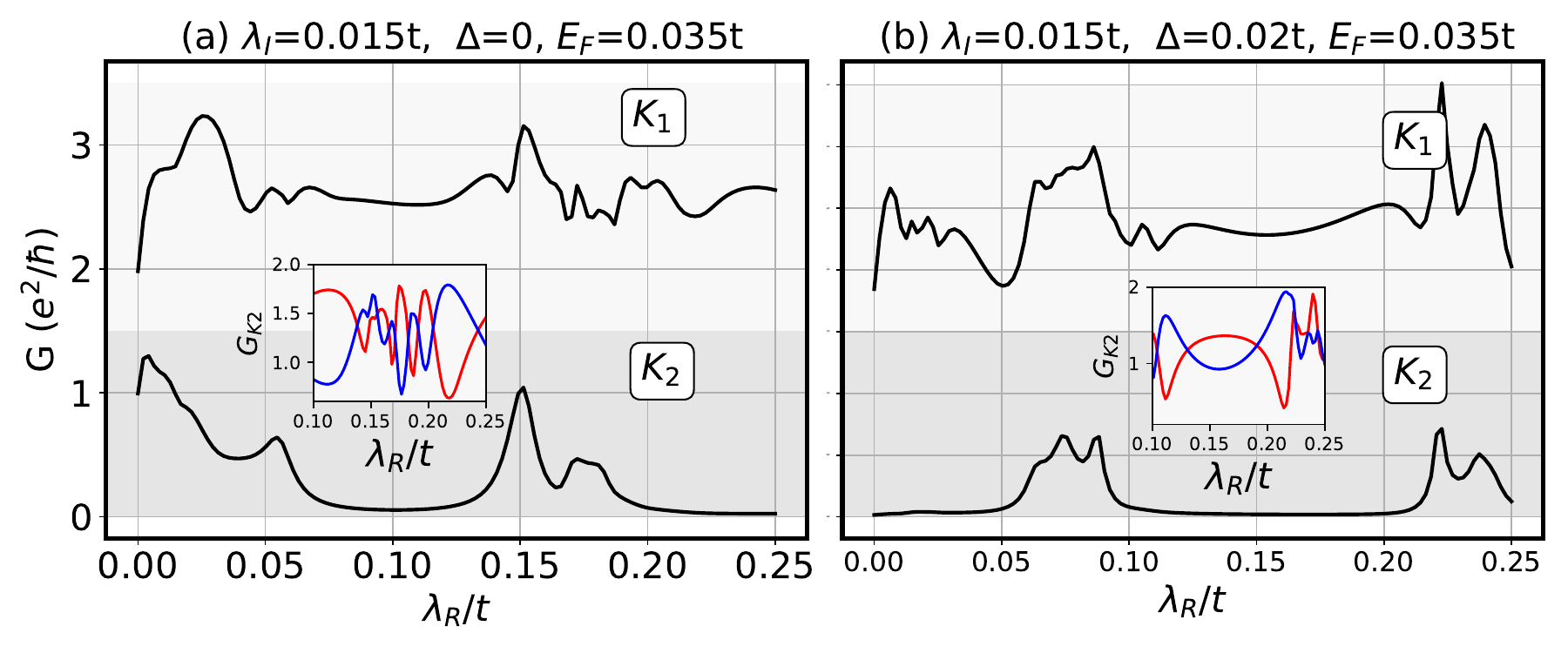}
	\includegraphics[width=9cm, height=4cm]{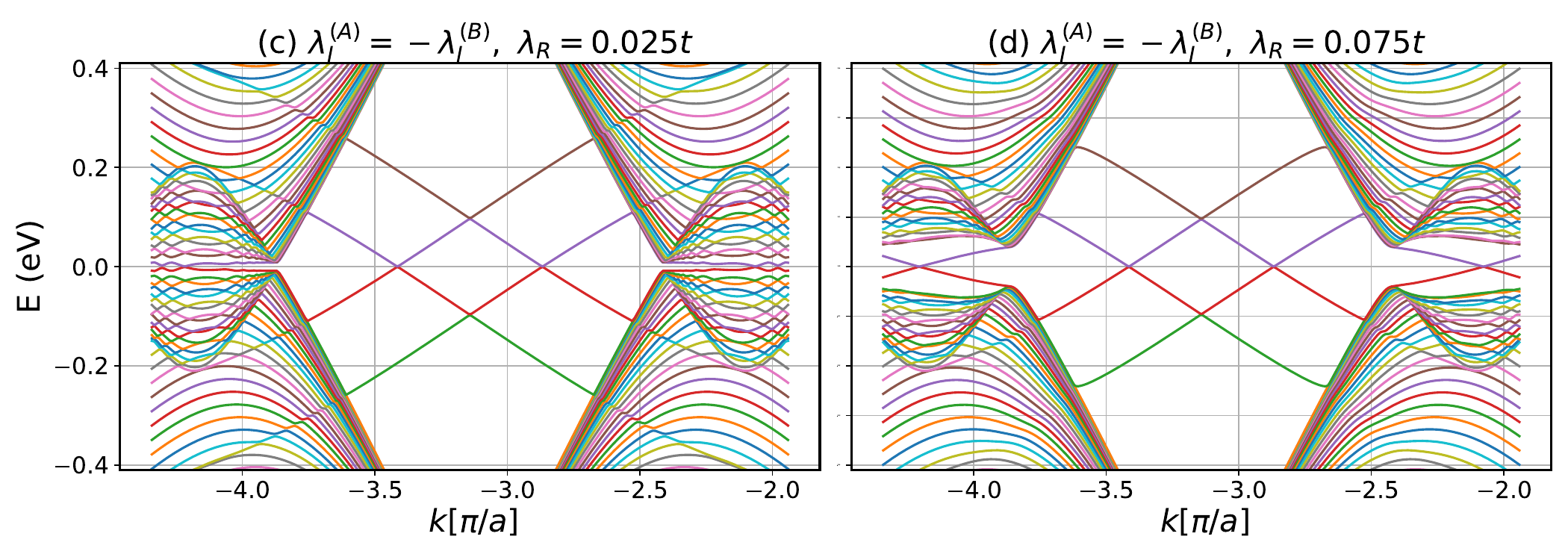}	
	\includegraphics[width=9cm, height=4cm]{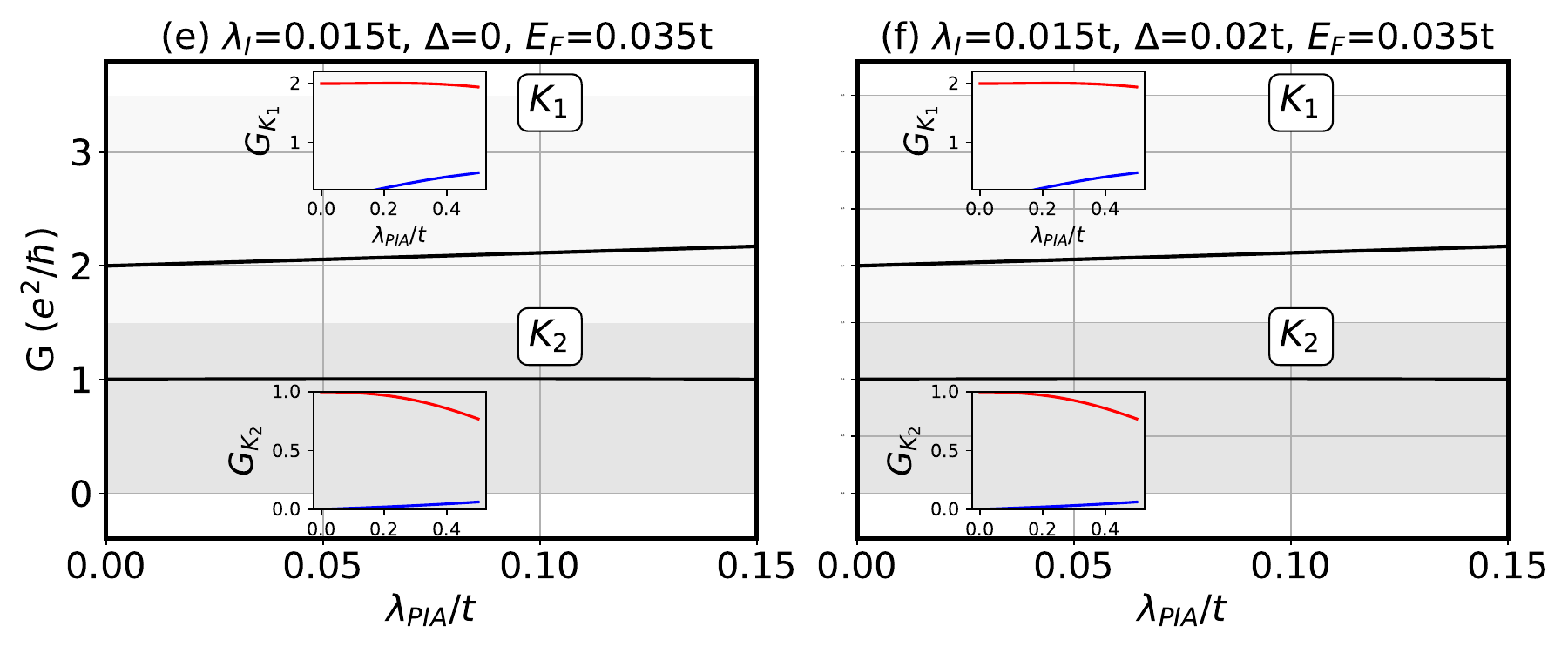}
	\includegraphics[width=9cm, height=4cm]{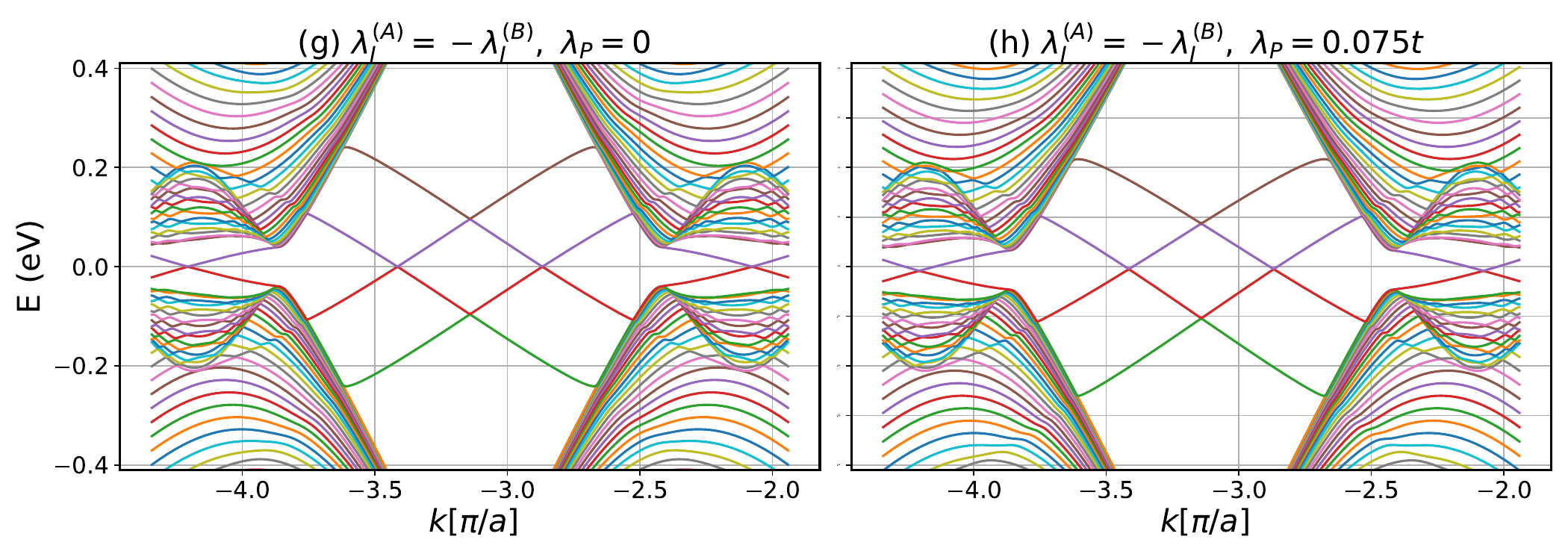}
	\vspace{-0.75cm}
	\caption{Panels (a) and (b): Valley conductance vs. Rashba spin-orbit coupling $\lambda_R$ for $\Delta=0$ and $\Delta=0.02t$, respectively. Panel (c) and (d) show the energy bands in the staggered case ($\lambda_I^{(A)}=-\lambda_I^{(B)}$) for $\lambda_R=0.025t$ and $\lambda_R=0.075t$, accordingly. Panels (e) and (f)  show the conductance vs the strength of the PIA coupling. Panel (g) and (h) show the energy bands for $\lambda_P=0$ and $\lambda_P=0.075t$, accordingly. Insets in (a, b, e, f) show spin-resolved valley conductance per valley.}
	\label{fig6}
\end{figure}

In Fig. \ref{fig6} (a) and (b), we show results for valley transmittance when the top gate is used to obtain a valley polarized conductance for coupling within the ranges $0.07t \le \lambda_R \le 0.12t$ and $0.1t \le \lambda_R \le 0.2t$,  for the two values of the staggered potential $\Delta=0$ anc $\Delta=0.02t$ correspondingly. The valley transmission for ${\bf K_1}$, $({\bf K_2})$ jumps almost from 2 to 0 for positive (negative) incident energy, indicating that only electrons with either valley ${\bf K_1}$ or valley ${\bf K_2}$ go through the system.

The most important conclusion from the results is that reasonable and better control of the valley transmission can be obtained by tuning a gate bias using IQDs with weak couplings.  In this context, a quantitative understanding might easily be reached. By controlling the Rashba strength, the wave function may interfere destructively or constructively depending on the valley and the sign of the incident energy. It is seen that at $E_F>0$ and $\lambda_R \ge 0.05t$, the wave function around ${\bf K_1}$ is transmitted through the IQDs region, where the wave function around ${\bf K_2}$  vanishes (is reflected), resulting on a polarized valley transmission, invertible by changes in the sign of the incident energy. 

These results can be understood from the analysis of the band structure in Fig. \ref{fig6} (c) and (d). By analyzing the RSOC's contributions to the edge state, we observe that for weak Rashba couplings, such as $\lambda_R = 0.025 t$, the edge state is mixed with the bulk conduction bands. However, for $\lambda_R = 0.075 t$, the system might sustain isolated and steady helical edge states, depending on the spin nature, as discussed in earlier works \cite{Belayadi-tmd-2023}.

%%---------------------------- Fig. 6--------------------
%%------------------------- end Fig. 6--------------------

In addition to the valley filtering, the interplay between the valley and spin degrees of freedom might realize new transport regimes. As shown in the inset of Fig.\ref{fig6} (a) and (b), the wave functions around ${\bf K_1}$ or ${\bf K_2}$ (depending on the incident energy) are spin-dependent. Indeed, the Rashba coupling allows off-diagonal spin-flipping terms. Therefore, the spin conductance is also directly controlled by the Rashba strength, leading to valley-spin interplay dynamics. For a given incident energy, the spin-conductance $G_{\uparrow}({\bf K_1})$ oscillations have several phase shifts that correspond to the wavelengths of the oscillations determined by the Rashba coupling strength. Notice that the overall conductance of spin-polarized carriers, $G_{\uparrow}({\bf K_1}) =G_{\uparrow \uparrow}({\bf K_1})+G_{\uparrow \downarrow}({\bf K_1})$ remains constant. Our calculations suggest valley processes dominate the conductance as spin-flips between neighboring sites require wider IQDs. Details on spin-dependent conductance in these proximity effect structures have been fully addressed in Ref. \cite{Belayadi-tmd-2023}.

\subsubsection{ IQDs with lower- or higher-order symmetries}\label{sub-b-2}
$C_{3v}$ symmetry is a property of hexagonal systems described by point group symmetry methods. To generalize the applicability of the model presented, it might be useful to lower (or raise) the symmetry of the IQDs. This strategy provides more options to describe novel regimes as it allows for removing (or adding) different spin-orbit terms. (i) In the case $\lambda_R=0$, the point group is increased to $D_{3h}$ where the sublattice inversion asymmetry defines a point group of a planar system sustaining a triangular lattice with two staggered (non-equivalent) sublattices  \cite{Kochan-2017}. For such a case, the presence of the ISOC leads to the transmittance spectrum similar to Fig \ref{fig2} (c) and (d).  (ii) To lower the symmetry of the IQDs, one might include an additional staggered SOC term called PIA coupling (pseudospin inversion asymmetry for short). The spin-orbit Hamiltonian, in this case, will have an extra term:
\begin{eqnarray}
H_{PIA} &=& (\sqrt3 a_0/2)\big[ \lambda_{PIA}^{(A)}( \sigma_z+\sigma_0) \\
        &+ &\lambda_{PIA}^{(B)}( \sigma_z-\sigma_0)\big] (k_x \sigma_y + k_y \sigma_x). 
\end{eqnarray}
where $\lambda_{PIA}^{(A)}$ and $\lambda_{PIA}^{(B)}$ are the staggered PIA coupling and $a_0$ is the lattice constant.

As shown in Fig. \ref{fig6} (e) and (f), the PIA coupling does not affect the valley-conductance (even when it is a staggered spin-orbit term), neither for lower nor higher strength values. Therefore, we can confirm that only ISOC and RSOC control the valley process, as discussed in Fig. \ref{fig2} and \ref{fig6} (a, b). Hence, a QD with $\lambda_{PIA} \ne 0$ brings forth the same behavior as $\lambda_{PIA} = 0$, and a similar valley response will be observed. Indeed, the spectrum of the energy bands in Fig. \ref{fig6} (g) and (h) supports these results, where we observe that for either neglected ($\lambda_{PIA} = 0$) or strong PIA terms ($\lambda_{PIA} = 0.075t$) the lower bulk and edge states are insensitive to the coupling strength.

This result is exciting as it implies that one might use TMDs to create IQDs with  $C_{3v}$-symmetry but with a $\lambda_{PIA} \ne 0$ without affecting their valley transport properties. Indeed, we might easily tune the intrinsic and Rashba parameters in such a system using TMD quantum dots. For instance, based on density functional theory fittings, both parameters are strongly sensitive to the electric field, twisting, and/or vertical strain effects \cite{Gmitra-2016, twisting-1, twisting-2}. The possibility of controlling the decrease or increase in the coupling strengths is highly desirable to monitor the valley polarization. Additionally, the spin-conductance is weakly affected (see inset of Fig. \ref{fig6} (c, d)) by the presence of PIA as compared to Rashba terms. For more details about spin dependence and the effect of PIA coupling on spin polarization, see \cite{Belayadi-tmd-2023-2}.

\subsubsection{Valley confinement with IQDs}
\label{sub-b-3}

This subsection addresses the possibility of attaining confinement of valley-polarized electrons. Adopting the same sample parameters as in subsection \ref{sub-a}, we see that by zooming in on the transmission spectra from Fig. \ref{fig2} (c, d), one can additionally recognize a resonance in the transmittance emerging in the IQDS, at $E_F=0.035 t $ for higher values of ISOC ($\lambda_{I}^{(A)}=-\lambda_{I}^{(B)}=0.065t$). The scattering through the induced region creates electron confinement for small-size IQDs ($r_0=7$ nm). The resonance might be advantageously produced for weak ISOC at specific incident energy values. This case defines a real proximity effect where we tune $E_F$ to confine the valley states as shown in Sec.\ref{sub-c}.

%%---------------------------- Fig. 7--------------------
\begin{figure}[tp]
	\centering
	\includegraphics[width=9cm, height=2.85cm]{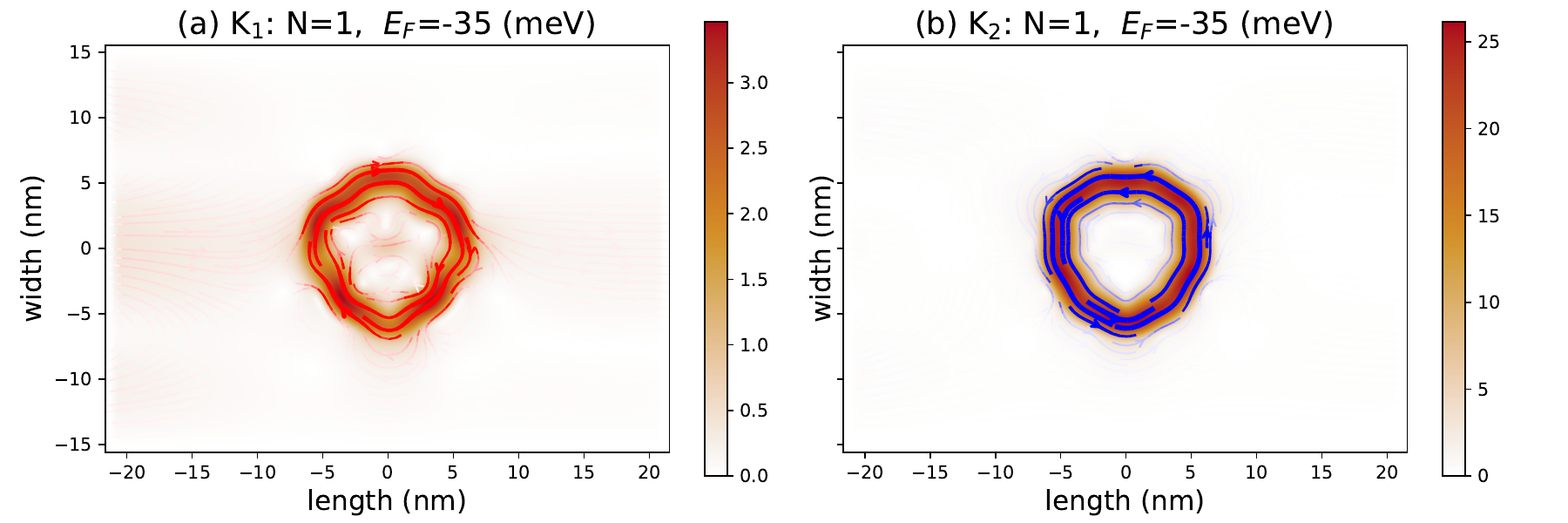}
	\includegraphics[width=9cm, height=2.6cm]{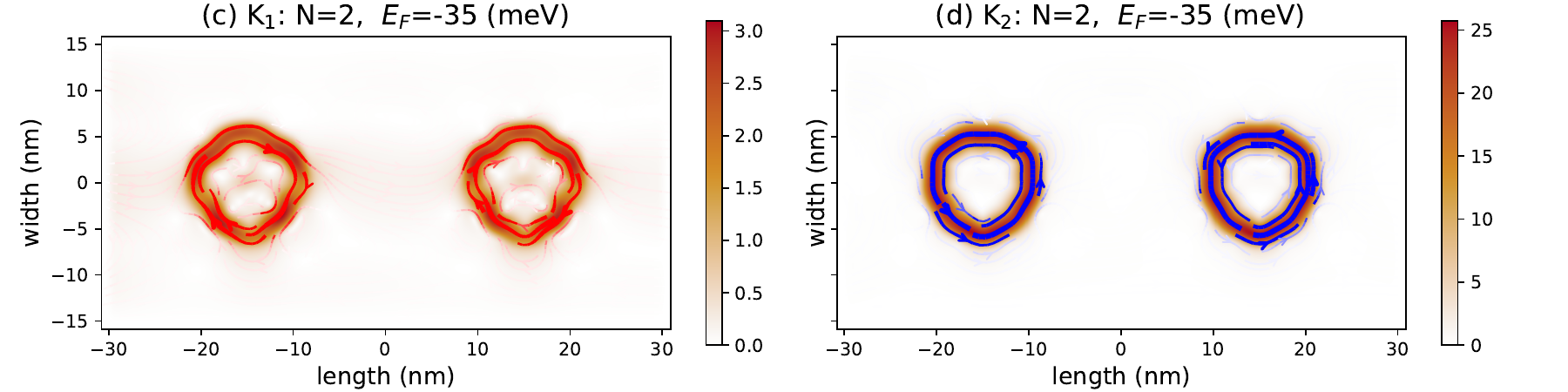}
	\vspace{-0.45cm}
	\caption{Local density of  valley-localized states near ${\bf K_1}$ (left) and ${\bf K_2}$ (right) at $E_F = 0.035t$, obtained for values of  $\lambda_{I}^{(A)}=-\lambda_{I}^{(B)}=0.065t$. The red and blue arrows show each valley state's circulation direction. Panels (a, b) and (c, d) are for single and 2-chain-IQDs.}
	\label{fig7}
\end{figure}

A detailed analysis of the local current profiles of valley states, as depicted in Fig. \ref{fig7}, shows a better representation of such a kind of valley-dependent electron confinement. Indeed, we observe that valley state confinement can be achieved by using either a single or a chain of IQDs. Importantly, the incident valley-states are trapped around the IQD region for appropriate SOC choices. This results in the confinement of the states, a process that can be regarded as the product of multiple internal reflections that trap the Dirac fermions by interference processes.

The system with symmetric IQDs appears to provide confinement for states at both valleys simultaneously in the same spatial region (we do not observe splittings of the valley confinement states). However, to achieve control over the valleys and multiple valleytronic and optoelectronic functionalities, using these IQDs, it will be necessary to create splittings and confining with unique quasi-bound states (belonging to either ${\bf K_1}$ or ${\bf K_2}$) in the local regions. The valley-bound states might be split by considering asymmetric quantum dots with different shapes and point group symmetries. This topic will be addressed in future work.

\subsection{Realistic IQDs in heterostructures of twisted graphene and TMD monolayers}
\label{sub-c}

As discussed in Sec. \ref{sub-b-2} and \ref{sub-b-3}, a quantitative study of the model has been addressed in the case of weak SOCs. To provide a comprehensive picture of concrete setups, we perform a real case study and show how an IQD might be realized in realistic experimental conditions. 

\begin{figure}[tp]
	\centering
	\includegraphics[width=9cm, height=4cm]{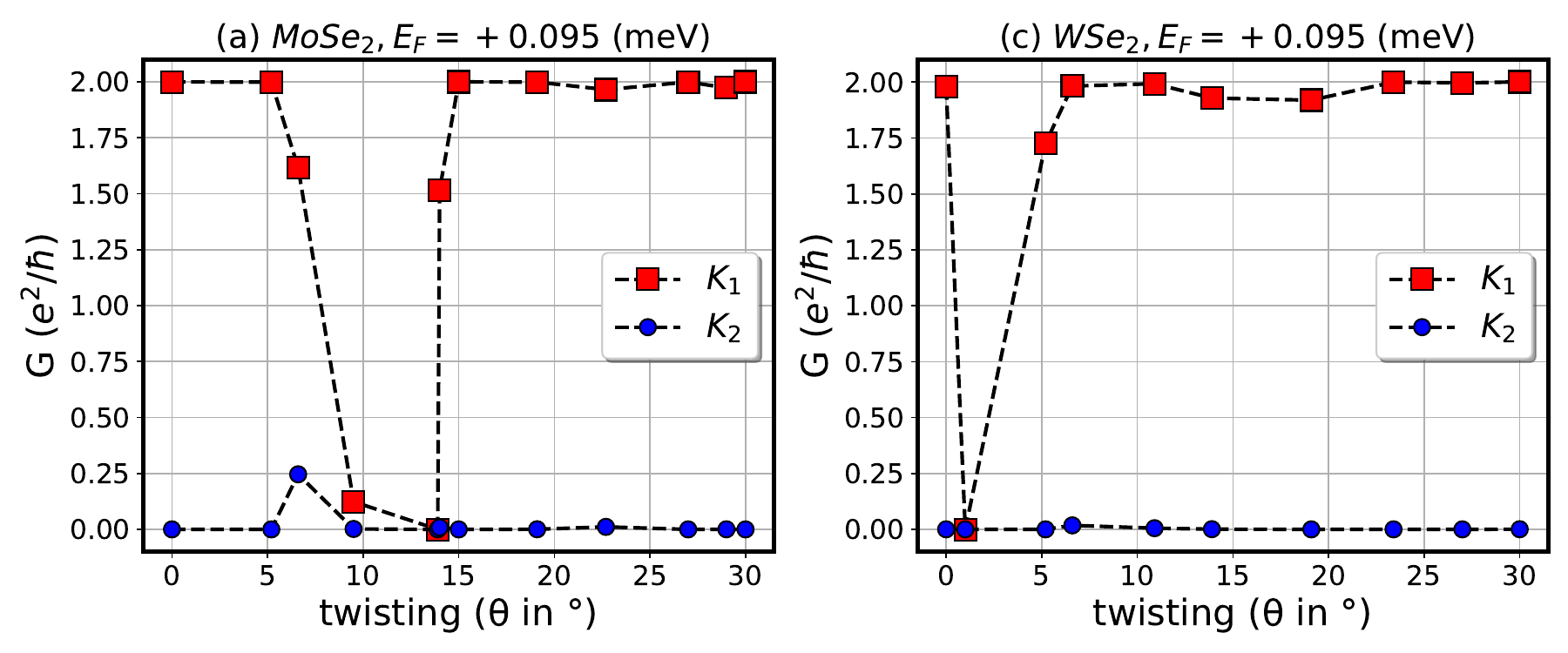}
	\includegraphics[width=9cm, height=4cm]{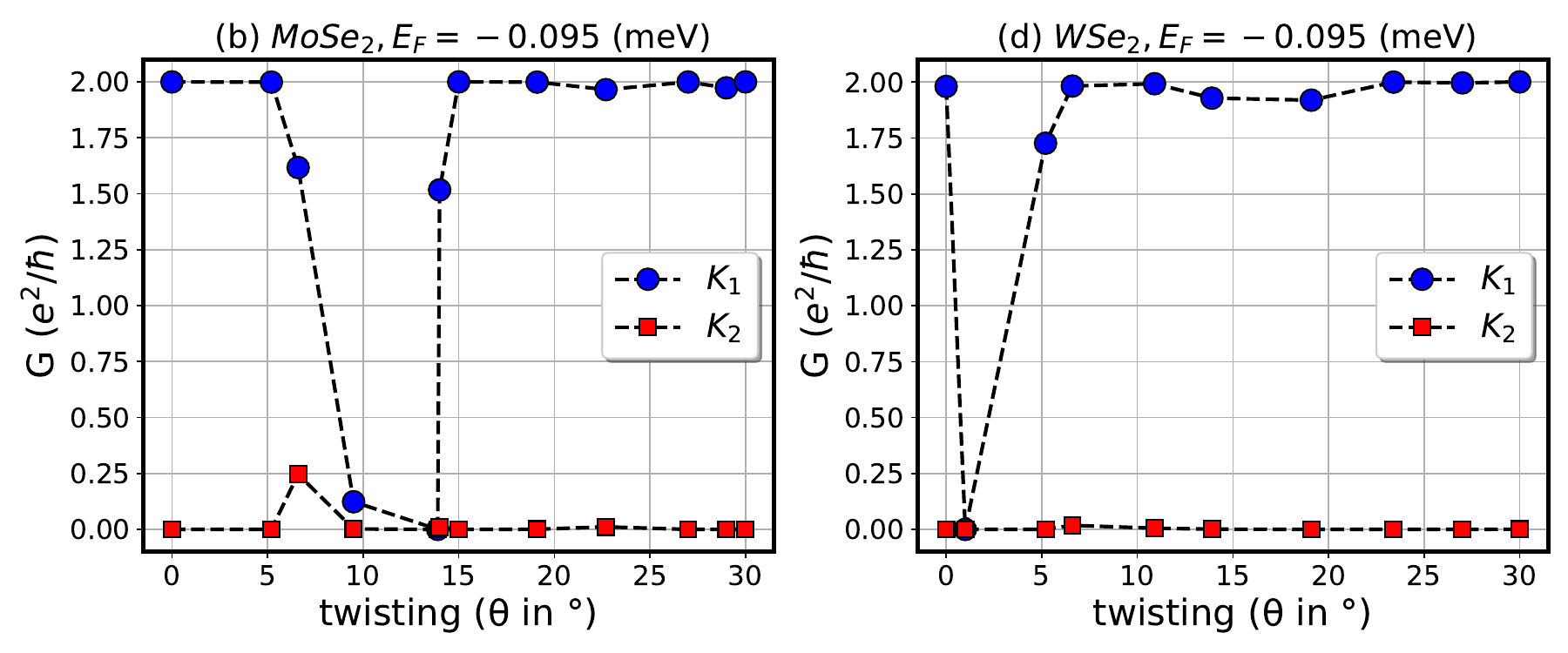}
	\includegraphics[width=9.4cm, height=2.5cm]{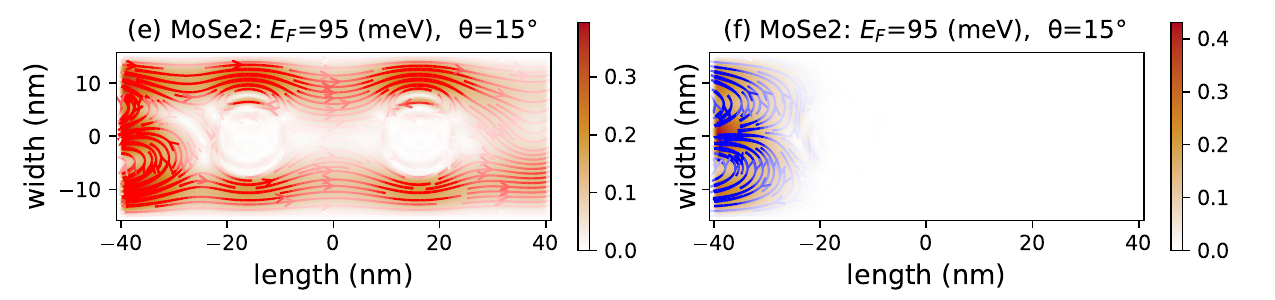}
	\includegraphics[width=9.4cm, height=2.5cm]{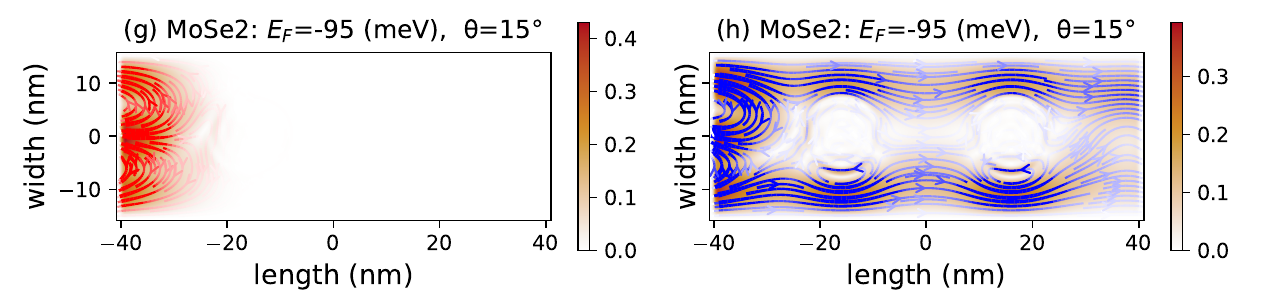}
	\vspace{-0.45cm}
	\caption{Panels (a) and (b) show valley conductance vs twist angle $\theta$ for IQDs made of MoSe$_2$ on graphene at $E_F = 0.095$meV and $-0.095$meV, respectively. Panels (c) and (d) correspond to IQDs of WSe$_2$ on graphene for the same values of $E_F$. In panels (e)-(f) and (g)-(h) we show valley-polarized currents for MoSe$_2$ for positive and negative bias ($E_F$), respectively}.
	\label{fig8}
\end{figure}

To start, we consider IQDs produced with four different semiconducting TMDs islands: MoSe$_2$, WSe$_2$, MoS$_2$, and WS$_2$ deposited on graphene monolayer, as shown in Fig. \ref{fig1} (b). Furthermore, we analyze the proximity-induced spin-orbit couplings in the case of twisted TMDs on graphene. Based on the work of T. Naimer et {\ et al.} \cite{Naimer-2021}, twisting leads to tuning the magnitudes of the valley-Zeeman and Rashba spin-orbit couplings. Interestingly, the amplitude of the staggered potential $\Delta$ in twisted G-TMDs is significantly weak and tends to zero at some specific twisting angles, thus favoring valley filtering, as discussed in Fig. \ref{fig4}. Based on the tight-binding parameters derived from first principles calculations \cite{Naimer-2021}, we compute the valley conductance through four $(n=4)$ IQDs at $E_F = 0.035 t \simeq0.095$meV and show results in Fig. \ref{fig8}.

%%---------------------------- Fig. 8--------------------

{\bf Valley transmittance with first propagating modes in twisted TMDs:} It is important to state that in all previous sections, we considered the transmittance spectrum at $E_F= \pm 0.035t$; a value that ensures that the valley-dependent conductance is addressed independently with the simultaneous excitation of the first propagating modes in ${\bf K_1}$ and ${\bf K_2}$ since the Fermi energy $E_F$ is larger than the valley-mode spacing gap. For more details, refer to our previous work \cite{belayadi-prb1}. 

In Fig. \ref{fig8}, we inject electrons with $E_F =\pm 0.095$meV into four identical IQDs produced with MoSe$_2$ and WSe$_2$ islands. The induced SOCs are tuned by twisting the TMDs and allow for monitoring of the valley response in the system.  At positive incident energy, the valley conductance from valley ${\bf K_1}, ({\bf K_2})$ is 2 (0) in the units of $e^2/ \hbar$. Only electrons from either valley ${\bf K_1}$ or ${\bf K_2}$  go through the system (see panels (a) and (c)). As previously discussed, we also confirm that changing the Fermi energy sign reverses the valley polarity, as shown in Fig. \ref{fig8} (b) and (d).
Additionally, as shown in  \ref{fig8} (a) and (b), twisting in the case of  MoSe$_2$ (that induces a weak staggered potential \cite{Naimer-2021}) leads to control of valley transmittances similar to what we found when discussing the model with weak couplings in Fig. \ref{fig6}.  

The obtained results with MoSe$_2$ can be applied to obtain pure valley polarized conductances with the twist angle outside the limits of $9^\circ - 14^\circ $. Hence, the valley transmission from ${\bf K_1}$ (${\bf K_2}$) jumps almost from 2 to 0 for positive (negative) incident energy, depending on twisting angles and full valley polarized currents. To emphasize these findings, we plot the local current for $\theta = 15^\circ $ at $E_F = 0.095$meV ($-0.095$meV) for MoSe$_2$ (WSe$_2$). Indeed, one observes that only one valley is scattered, which leads to conducting Hall currents depending on the sign of the incident Fermi-energy. These results are important compared to those shown in Fig. \ref{fig5}, where both valleys conduct through bulk and Hall currents. Hence, the IQDs based on semiconducting TMDs MoSe$_2$ and WSe$_2$ might be used as promising islands for generating valley Hall signals.

%%---------------------------- Fig. 9--------------------
\begin{figure*}[tp]
	\centering
	%\hspace*{-0.4cm}
	\includegraphics[width=14cm, height=4cm]{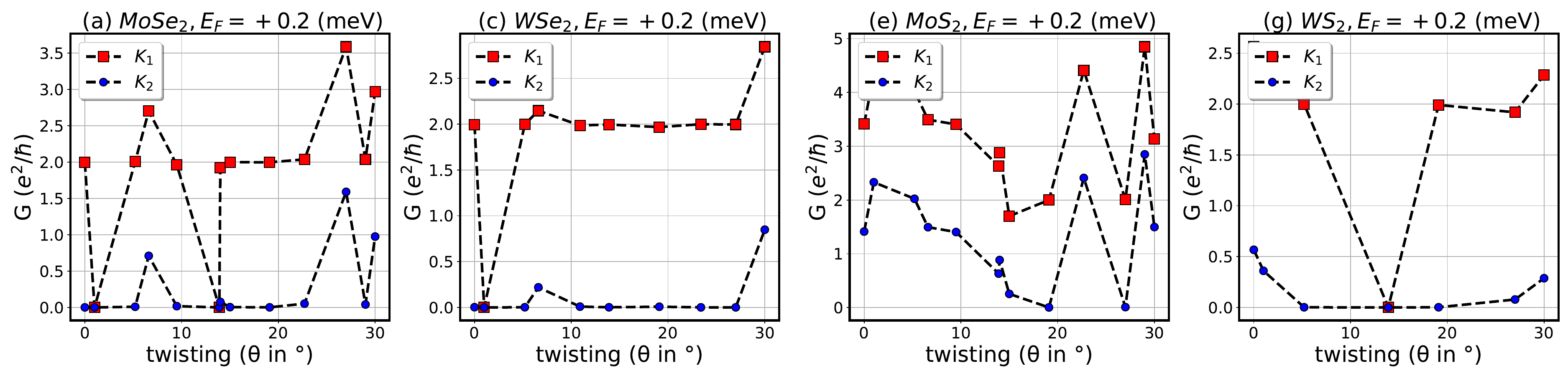}
	\includegraphics[width=14cm, height=4cm]{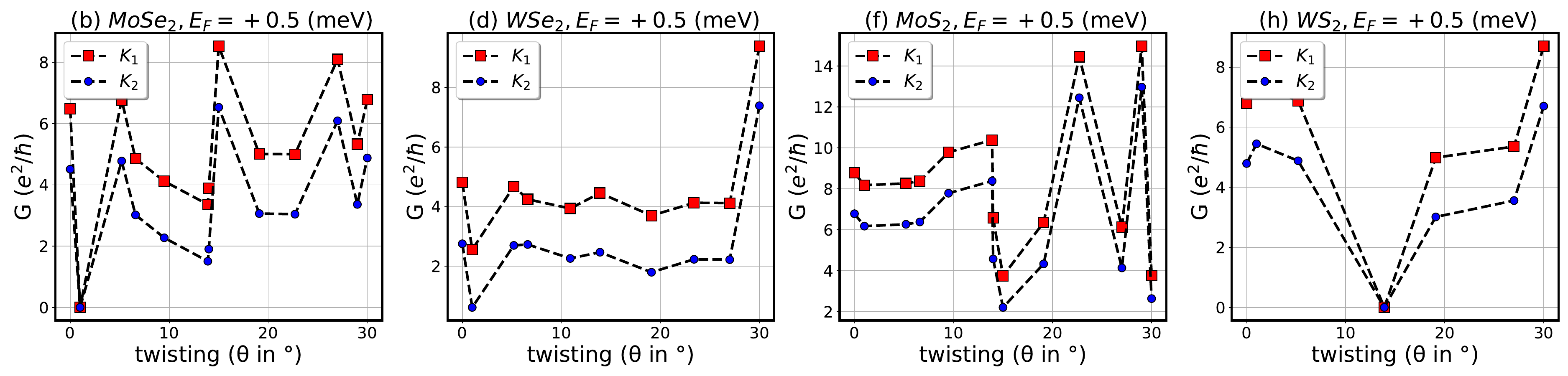}
	\vspace{-0.45cm}
	\caption{Valley conductance vs twist angle $\theta$ for IQDs made of MoSe$_2$, WSe$_2$, MoS$_2$ and WS$_2$ on graphene, respectively. Panels in the top (bottom) show conductances at $E_F=0.2$meV (0.5)meV.}
	\label{fig9}
\end{figure*}

{\bf Valley transmittance beyond first propagating modes in twisted TMDs:}
To observe transport involving states beyond the first propagating modes, we set the incident energy of the injected electrons at $E_F= \pm 0.1, \pm 0.3$meV. In this case, we will consider all TMDs: MoSe$_2$, WSe$_2$, MoS$_2$, and WS$_2$ where the valley-dependent conductance is shown in Fig. \ref{fig9}. Similarly to what was discussed, the valley process is sensitive to the magnitudes of the induced valley-Zeeman and Rashba spin-orbit couplings and the incident Fermi energy. The TMD islands, as well as the twisting, allow a variety of options (several values of ISOCs) to monitor valley-driven currents, either in bulk or along the edges, where the best choice is to set the incident energy at $E_F=0.035t$ (within the first propagating modes).
We observe that for some of the twist angle values, the transmittance of both valleys is zero, which might be explained either by zero current (OFF) or by the presence of the confinement states.

\begin{figure}[tp]
	\centering
	\includegraphics[width=9cm, height=2.5cm]{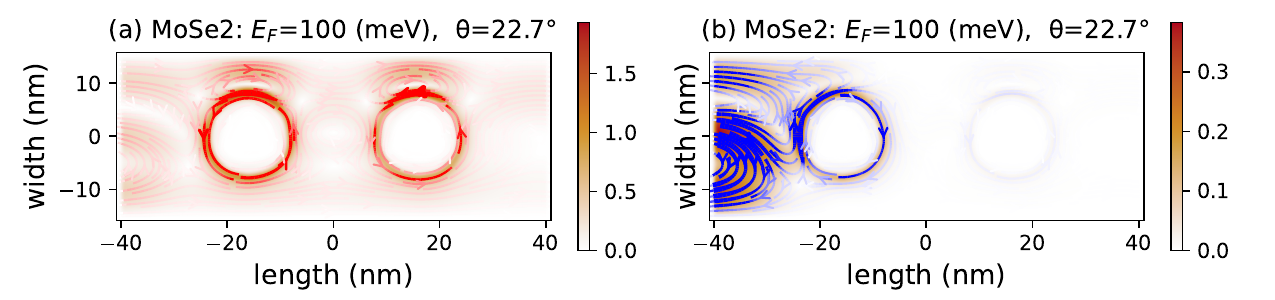}
	\includegraphics[width=9cm, height=2.5cm]{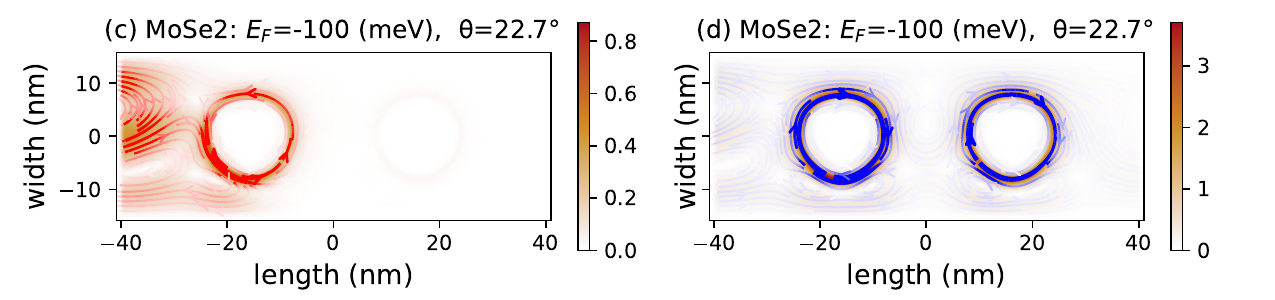}
	\includegraphics[width=9cm, height=2.5cm]{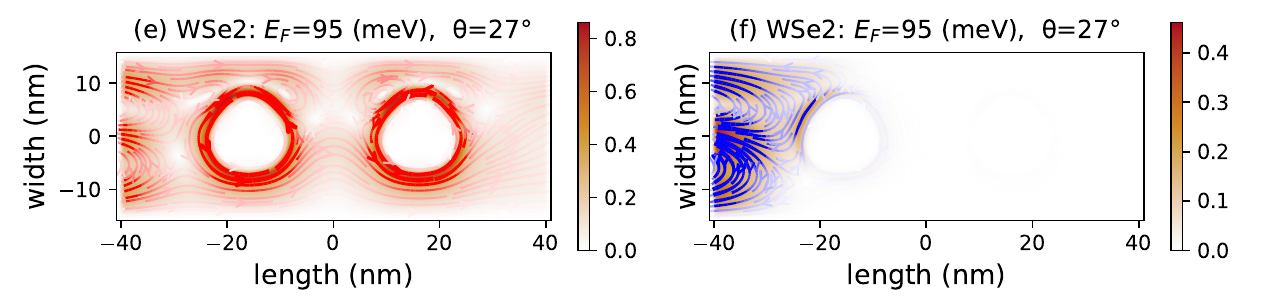}
	\includegraphics[width=9cm, height=2.5cm]{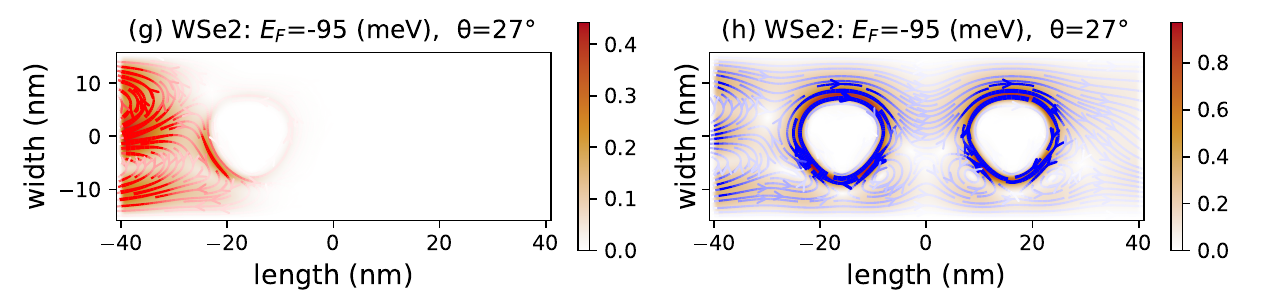}
	\includegraphics[width=9cm, height=2.5cm]{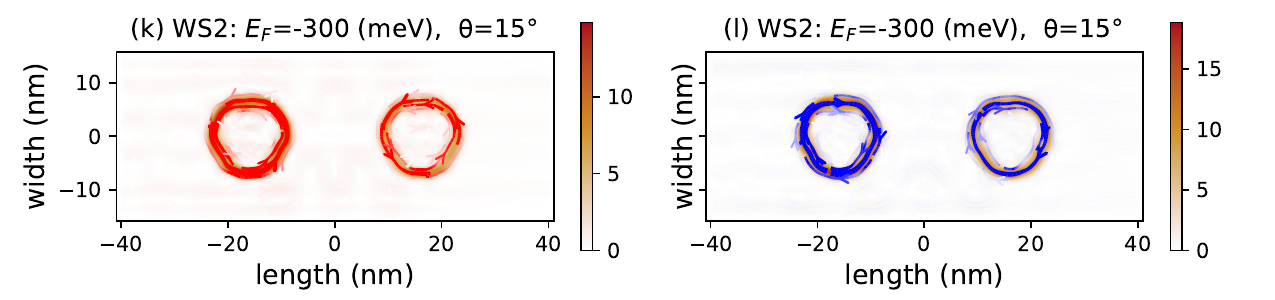} 
	\vspace{-0.45cm}
	\caption{Real-space profiles for both valley currents: red (blue) lines show the current profile from states originating in valley ${\bf K_1} \quad ({\bf K_2}$). Panels (a)-(d) show the valley conductance vs twist angle $\theta$ for IQDs made of MoSe$_2$ on graphene at $E_F=\pm 100$ meV. Panels (e)-(h) are for WSe$_2$ on graphene at $E_F=\pm 95$ meV. Panels in (k) and (l) correspond to WS$_2$ at higher energy $E_F=-300$ meV.}
\label{fig10}
\end{figure}

{\bf Valley confinement in twisted TMDs:}
Previously, we have shown in Fig. \ref{fig7} that at $E_F = 0.035t$, the resonance or confinement might be produced at higher values of ISOC strengths. This feature is somehow challenging to attain in available setups. To confirm the presence of resonances in realistic conditions, we tune both the incident Fermi energy and the twist angle. In Fig. \ref{fig10}, we show that the resonance condition might be obtained at weak induced SOCs at some specific incident energy and twist values. Importantly, we observe that at $\theta = 22.7^{\circ}$ ($27^{\circ}$), the system can confine valley states in the IQD region of MoSe$_2$ (WSe$_2$) at a lower energy $E_F=\pm 0.035t=\pm 0.098$meV (within the range of the first propagating mode). Indeed, for appropriate choices of the SOC values related to the chosen TMD and respective twist angle, the incoming electron is trapped around the IQD area, as shown in Fig. \ref{fig10}. More precisely, for $E_F>0$, the IQDs of MoSe$_2$ and WSe$_2$ islands (panels Fig. \ref{fig10} (a), \ref{fig10} (b), \ref{fig10} (e), \ref{fig10} (f)) reveal higher localization throughout the scattering regions, supported by valley-localized states from valley ${\bf K_1}$. However, the valley-localized states from ${\bf K_2}$ are blocked at the first IQD. The process might be reversed for $E_F<0$ as we have shown in the lower panels of Fig. \ref{fig10} \ref{fig10} (c), and \ref{fig10} (d),  for MoSe$_2$ and Fig. \ref{fig10} \ref{fig10} (g), and \ref{fig10} (h) for WSe$_2$. This is an important result since this class of materials allows or induces valley confinement that might be used to process the optical responses and detect and monitor valley polarization \cite{Krasnok, Sharma, Rasmus}.

By increasing the incident energy beyond the first propagating mode, it is possible to confine both valley states around the IQDs of WS$_2$ islands as shown in Fig. \ref{fig10} (k) and (l). Both confined states are highly localized around all IQDs in the scattering region, as predicted in \cite{Sie}. Interestingly, shifting the sign of the incident energy reverses the path of interference, and both valleys exchange the propagating direction.

%%---------------------------- Fig. 10--------------------

\section{Conclusions and summary }\label{sec4}
We have used a tight-binding approach to obtain an effective, low-energy Hamiltonian to investigate valley-dependent transport through an array of proximity-induced quantum dots with $C_{3v}$ symmetry in a zigzag graphene nanoribbon. The model is a natural extension of the structure studied in Ref.\cite{Ren-2023} that analyzed spin and charge distributions in a graphene quantum dot induced by a WSe$_2$ island. Because the fabrication of graphene/TMD heterostructures with quantum dots of one or the other material is rapidly evolving \cite{Manikandan-2019, Singh-2021, Alfieri-2022}, experimental implementations for arrays of quantum dots will likely be realized in the near future. 

Our results show that the valley conductance exhibits an interesting behavior sensitive to the model parameters and the values of the various SOCs induced by the islands that decorate the ribbon. At the same time, this sensitivity is a great advantage for monitoring and controlling the different valley filter regimes observed. The results reveal the localization of valley-centered states produced by the competition between Zeeman and Rashba couplings in narrow IQDs. For larger IQDs, the valley states combine with bulk states, and the valley polarization is considerably deteriorated \cite{Frank-2018}.

Following Ref. \cite{Kane-Mele}, the qualitative description of the model shows that by varying Zeeman couplings, the conductance through a symmetric chain of quantum dots displays square-shaped curves with wide gaps. However, these features tend to vanish for some large ISOC values with the subsequent vanishing of valley polarization. Furthermore, for specific ranges of ISOC values, both valleys are present, implying that the device does not display a perfect valley-transistor behavior. An important conclusion from
 the qualitative description is the similarity of this system to the Datta-Das transistor. In both cases, the spin conductance is directly controlled by the strength of the Rashba SOC. In this respect, the most important SOC for good valley polarization is the staggered intrinsic SOC. These features could be helpful for further exploration of actual devices.

The sensitivity to the Rashba coupling is discussed in the presence of weak ISOC to describe realistic settings where proximity effects develop. The valley conductance shows that tuning the valley polarization and switching the valley scattering in the system using a top gate is possible. The possibility of using TMDs as decorating islands to form IQDs has also been discussed. We have shown that the presence of the PIA coupling, characteristic of these structures, does not affect the valley-polarization. Therefore, such islands can be used to tune the polarization by either strain or twists since the Rashba and Zeeman coupling are sensitive to external electric or strain field effects \cite{Gmitra-2016, twisting-1, twisting-2}.

Finally, we have applied these models to solve examples of realistic material combinations. To give a comprehensive picture, we have used graphene/TMD heterostructures with different semiconducting materials: MoS$_2$, MoSe$_2$, WSe$_2$, and WS$_2$. We noticed that IQDs based on semiconducting TMDs might be used as promising islands for generating valley Hall signals. Indeed, TMDs allow valley filtering processes and break the valley degeneracy, producing a valley-polarized current that favors valley selection by tuning the sign of the incident Fermi energy or the value of twist angles. Notably, the TMD island and control of the twisting angle allow various options (as they determine different values of ISOCs) to monitor valley-driven currents, confine both valley states simultaneously in the same region, or split the valley confinement states. 

Achieving a nearly square-wave transmission and a valley-valve effect for the ${\bf K_1}$ or ${\bf K_2}$ valleys is highly desirable for device applications. Moreover, the confinement of quasi-bound states from either valley is extremely important for manipulating optoelectronic interactions \cite{Zhang, Aktor, Steven} and valley-qubit systems  \cite{Wu,Szechenyi,Pawlowski}. Furthermore,  the mechanism for generating valley-Hall conductivity with valley-neutral currents could be handy to obtain pure valley-Hall signals. The system proposed in this manuscript exhibits features close to these goals; however, asymmetric QDS with controlled shape and several induced point group symmetries could lead to richer results. We plan to address this issue in future work.

%% *************************************
\acknowledgments
%\vspace{-0.5cm}
The authors acknowledge computing time on the SHAHHEN supercomputers at KAUST University, Saudi Arabia, and the supercomputers at the Centre for Research in Molecular Modeling (CERMM), Richard J. Renaud Science Complex, Concordia University.
AB would like to thank Dr. Adel Abbout at KFUPM, Saudi Arabia, for helpful discussions.
\clearpage
%\ \\
%\  \\	
%	
%%----------------------------------------------------------------------------------------------------
%%%%%%%%%%%%%%%%%%%%%%%%%%%%%%%%%%%%%%%%%%
%\vspace{6pt}
%%%%%%%%%%%%%%%%%%%%%%%%%%%%%%%%%%%%%%%%%%
%%%%%%%%%%%%%%%%%%%%%%%%%%%%%%%%%%%%%%%%%%
	
\clearpage
\appendix
\numberwithin{equation}{section}
\begin{widetext}
\begin{appendices}

\section{Valley-dependent conductance}\label{app-A}
As stated, we use the scattering matrix formalism to calculate the individual valley conductances. 
We separate the propagating modes in the leads depending on their velocity and momentum direction using the Kwant package \cite{kwant}. 
In the scattering region, valley currents might be mixed due to inter-valley scattering. However, as the two valleys are far apart in the Brillouin zone, effective valley mixing will require short-range potentials. A graphene membrane wraps smoothly for graphene deposited on top of islands to minimize strain effects. Thus, the sharp atomic termination of the island (that could give rise to inter-valley mixing) is effectively masked. These issues will have a negligible contribution for islands on top of graphene if the islands and their separation are large enough \cite{J.G.Nedell}.

To this end, we consider only propagating states for which $\Phi({\bf vF} > 0)$. The applied source-drain voltage chooses the direction of the current composed of these states, which have both spin and valley degrees of freedom. We focus first on the valley degree and lift the valley degeneracy by defining the propagating wave functions $\Phi_{\bf K_1} = \Phi({\bf k} > 0), \Phi_{\bf K_2} = \Phi({\bf k} < 0)$. 

These wave functions independently solve the scattering problem in the reciprocal space. To implement the Green's function formalism \cite{Datta-1995} we need to include the scattering matrices $S^{mn}= S_{\bf K_1}^{mn}+ S_{\bf K_2}^{mn}$ with 
$S_{\bf K_{1,2}}^{mn}$ given by
\begin{equation}\label{eqA1}
S_{\bf K_{1,2}}^{m, n}= \text{Tr} [G_{\bf K_{1,2}} \Gamma^m \, G^{\dagger}_{\bf K_{1,2}} \, \Gamma^n], \qquad 
(m, n = L, R \ or \ R, L);
\end{equation}
The Green's function and $\Gamma$ matrices are given by
\begin{eqnarray}
\label{eqA2}
G (\epsilon, {\bf K_{1,2}}) &=& \left[ \left( \epsilon + i \eta \right)I-H_{QD}({\bf K_{1,2}})- \Sigma \right]^{-1}\\ 
\Gamma &=& i (\Sigma - \Sigma^{\dagger}).  
\end{eqnarray}
$\Gamma$ defines the self-energy of the contacts placed to the left and right of the scattering region, and the relevant  Hamiltonian is $H_{QD}$, cf. Eq. (\ref{eq1}). Then, for each valley mode, the valley conductance at the Dirac cones is given by Eq. (\ref{eq2}).

%==================================================
\section{\textit{Valley-resolved current}}\label{app-B}

To obtain the local density of states (LDOS) and currents per valley, we obtain the wave functions of the propagating modes $\Phi$ for a given energy $E$ and site $i$. The propagating wave functions are stored per site depending on their momentum $\left\{ \Phi({\bf K_1}), \ \Phi({\bf K_2}) \right\}$ and their spin degree of freedom. The resulting LDOS, at a given site $i$ in the sample, is defined by
\begin{equation}\label{eqA5}
\text{LDOS}^{\bf K_{1,2}}\left( E, i \right)=\sum_{l}^{}\left| %\left
\langle i|\Phi^{\bf K_{1,2}}_l\rangle \right|^{2}\delta (E-E_l )
\end{equation}
where the sum is over all electron eigenstates $|\Phi_ l\rangle = c_l^{\dagger}|0\rangle$ of the Hamiltonian $H_{QD}$ in Eq. (\ref{eq1}) with energy $E_l$. The valley-resolved LDOS in  Eq. (\ref{eqA5}) is calculated using Chebyshev polynomials \cite{af2015} and damping kernels \cite{Wellein}. 

The corresponding density operator and the continuity equation are expressed as
\begin{equation}\label{eqB1}
\rho_q^{\bf K_{1,2}}=\sum_{a}[\Phi_{a}^{\bf K_{1,2}}]^* \, H^{s}_{q} \, \Phi_{a}^{\bf K_{1,2}}, 
\qquad \frac{\partial \rho_a^{\bf K_{1,2}}}{\partial t}-\sum_{b} J_{a, b}^{\bf K_{1,2}}=0.
\end{equation}
where $a$ refers to all sites in the scattering region, $J_{ab}^{\bf K_{1,2}}$ is the valley-resolved current, and $H^s_q$  is the $q$-Fourier component of the hopping term in the scattering region.

For a given site density $\rho_a$, we sum over all the neighboring sites $b$. As a result, the valley current $J^{{\bf K_{1,2}}}_{ab}$ takes the form 
\begin{equation}\label{eqB2}
J_{a, b}^{\bf K_{1,2}}=[\Phi^{\bf K_{1,2}}({\bf v > 0})]^*\left(i \sum_{\gamma}^{}H^{*}_{ab\gamma} H^s_{a \gamma} - H^s_{a \gamma} H_{ab \gamma} \right) [\Phi^{\bf K_{1,2}}({\bf v> 0})]
\end{equation}
where $H_{a b \gamma}$ is a component of a rank-4 tensor that can be represented as a vector of matrices, and $\gamma$ is an index that runs over sites in real space. In this expression, Latin indices go over sites, and Greek indices run over the degrees of freedom in the Hilbert space. For more details about obtaining the current operator, see references \cite{belayadi-prb1, Belayadi-tmd-2023, Belayadi-tmd-2023-2, kwant}.
\end{appendices}
\end{widetext}

\end{document}